\documentstyle[psfig]{mn}

\newif\ifAMStwofonts
\ifoldfss
  \ifCUPmtlplainloaded \else
    \NewTextAlphabet{textbfit} {cmbxti10} {}
    \NewTextAlphabet{textbfss} {cmssbx10} {}
    \NewMathAlphabet{mathbfit} {cmbxti10} {} % for math mode
    \NewMathAlphabet{mathbfss} {cmssbx10} {} %  "   "    "
  \fi
  \ifAMStwofonts
    \ifCUPmtlplainloaded \else
      \NewSymbolFont{upmath} {eurm10}
      \NewSymbolFont{AMSa} {msam10}
      \NewMathSymbol{\upi}     {0}{upmath}{19}
      \NewMathSymbol{\umu}     {0}{upmath}{16}
      \NewMathSymbol{\upartial}{0}{upmath}{40}
      \NewMathSymbol{\leqslant}{3}{AMSa}{36}
      \NewMathSymbol{\geqslant}{3}{AMSa}{3E}

      \let\leq=\leqslant \let\le=\leqslant
      \let\geq=\geqslant 
    \fi
  \fi
\fi % End of OFSS

\ifnfssone
  \newmathalphabet{\mathit}
  \addtoversion{normal}{\mathit}{cmr}{m}{it}
  \addtoversion{bold}{\mathit}{cmr}{bx}{it}
  \newmathalphabet{\mathbfit} % math mode version of \textbfit{..}
  \addtoversion{normal}{\mathbfit}{cmr}{bx}{it}
  \addtoversion{bold}{\mathbfit}{cmr}{bx}{it}
  \newmathalphabet{\mathbfss} % math mode version of \textbfss{..}
  \addtoversion{normal}{\mathbfss}{cmss}{bx}{n}
  \addtoversion{bold}{\mathbfss}{cmss}{bx}{n}
  \ifAMStwofonts
    \ifCUPmtlplainloaded \else
      \UseAMStwoboldmath
      \makeatletter
      \new@mathgroup\upmath@group
      \define@mathgroup\mv@normal\upmath@group{eur}{m}{n}
      \define@mathgroup\mv@bold\upmath@group{eur}{b}{n}
      \edef\UPM{\hexnumber\upmath@group}
      \new@mathgroup\amsa@group
      \define@mathgroup\mv@normal\amsa@group{msa}{m}{n}
      \define@mathgroup\mv@bold\amsa@group{msa}{m}{n}
      \edef\AMSa{\hexnumber\amsa@group}
      \makeatother
      \mathchardef\upi="0\UPM19
      \mathchardef\umu="0\UPM16
      \mathchardef\upartial="0\UPM40
      \mathchardef\leqslant="3\AMSa36
      \mathchardef\geqslant="3\AMSa3E

      \let\leq=\leqslant \let\le=\leqslant
      \let\geq=\geqslant 
    \fi
  \fi
\fi % End of NFSS release 1

\ifnfsstwo
  \DeclareMathAlphabet{\mathbfit}{OT1}{cmr}{bx}{it}
  \SetMathAlphabet\mathbfit{bold}{OT1}{cmr}{bx}{it}
  \DeclareMathAlphabet{\mathbfss}{OT1}{cmss}{bx}{n}
  \SetMathAlphabet\mathbfss{bold}{OT1}{cmss}{bx}{n}
  \ifAMStwofonts
    \ifCUPmtlplainloaded \else
      \DeclareSymbolFont{UPM}{U}{eur}{m}{n}
      \SetSymbolFont{UPM}{bold}{U}{eur}{b}{n}
      \DeclareSymbolFont{AMSa}{U}{msa}{m}{n}
      \DeclareMathSymbol{\upi}{0}{UPM}{"19}
      \DeclareMathSymbol{\umu}{0}{UPM}{"16}
      \DeclareMathSymbol{\upartial}{0}{UPM}{"40}
      \DeclareMathSymbol{\leqslant}{3}{AMSa}{"36}
      \DeclareMathSymbol{\geqslant}{3}{AMSa}{"3E}

      \let\leq=\leqslant \let\le=\leqslant
      \let\geq=\geqslant 
    \fi
  \fi
\fi % End of NFSS release 2

\ifCUPmtlplainloaded \else
  \ifAMStwofonts \else % If no AMS fonts
    \def\upi{\pi}
    \def\umu{\mu}
    \def\upartial{\partial}
  \fi
\fi    

\title{The template type Ia supernova 1996X\thanks{Based on observations
 collected at ESO-La Silla.}}
\author[Salvo, M. E., et al.]{Salvo, M. E.$^1$, Cappellaro, E.$^1$, Mazzali, P. A.$^{2,3}$,
   Benetti, S.$^4$, Danziger, I. J.$^2$, \and Patat, F.$^5$, Turatto, M.$^1$
\\
$^1$Osservatorio Astronomico di Padova, Vicolo dell' Osservatorio, 5, I-35122 Padova, Italy\\
$^2$Osservatorio Astronomico di Trieste, Via Tiepolo, 11, I-34131 Trieste, Italy\\
$^3$Research Centre for the Early Universe, School of Science, University of Tokyo, Bunkyo-ku, Tokyo 113-0033, Japan \\  
$^4$Telescopio Nazionale ``Galileo'', Apartado de Correos 565, E-38700, Santa Cruz de La Palma, Canary Islands, Spain \\
$^5$European Southern Observatory, Karl-Schwarzschild Str. 2, D-85748, Garching Bei M{\"u}nchen, Germany} 
\date{Accepted .....
      Received ....;
      in original form ....}               

\pagerange{\pageref{firstpage}--\pageref{lastpage}}
\pubyear{1999}

\begin{document}

\maketitle

\label{firstpage}

\begin{abstract}    
UBVRIJ photometry and optical spectra of the Type Ia SN~1996X obtained
at ESO during a one-year-long observational campaign are presented,
and supplemented by late time HST photometry.  Spectroscopically,
SN~1996X appears to be a `normal' SN~Ia.  The apparent magnitude at
maximum was $B=13.24\pm0.02$, and the colour $B-V=0.00\pm0.03$. The
luminosity decline rate, $\Delta m_B(15) = 1.31\pm0.08$, is close to
average for a SN~Ia. The best estimate of the galactic extinction is
$A_B=0.30\pm0.05$, and there is evidence that reddening within the
parent galaxy is negligible.

Detailed comparison of the light and colour curves of various `normal' SNe~Ia
shows that the assumption that multicolour light curves can be described simply
as a one-parameter family is not perfect. Together with  problems in the
calibration of the templates, this may explain the discrepancies in the
distance modulus derived adopting different calibrations of the absolute
magnitude vs. light curve shape relations. Indeed we found that M$_B$ ranges
from $-19.08$ to $-19.48$ and $\mu$ range from 32.02 to 32.48 depending on the
method used \cite{hamuy_absmag,phil99,riess98a}.

Computations of model light curve and synthetic spectra for both
early- and late-times, confirm that 1996X is a normal Type Ia SN and
that a satisfactory fit can be obtained using a W7 progenitor
structure only if we adopt the short distance. A larger distance would
imply a too large Ni mass for this fainter than average SNIa.

\end{abstract}     


\section{Introduction}   

One of the most exciting and unexpected results in recent astronomy is
the apparent necessity for a finite cosmological constant, the result
of measuring cosmic distances using Type Ia Supernovae (SNe~Ia) as
distance indicators. This result relies heavily on the assumption that
the absolute magnitude of SNe~Ia can be accurately calibrated based on
the observed luminosity evolution
\cite{riess98a,perlm}.

The suggestion that the absolute magnitude of SNe~Ia correlates with
the luminosity decline rates dates back to Pskovskii (1967,
1977). However, only in the last decade, with the improvement in
photometric accuracy made possible by CCD detectors, has 
this correlation been quantified, although its calibration and
range of application are still debated (Phillips 1993; Phillips et
al.\ 1999; Hamuy et al.\ 1996a,b,c; Riess et al.\ 1996,
1998a).  Nearby SNe are essential for this approach because their
absolute luminosity can be calibrated accurately using other distance
indicators, e.g.  Cepheid variables \cite{sandage,madore} or the
Globular Cluster Luminosity Function \cite{mdv,dren}.

Theoretically, the effort to build a scenario consistently linking
progenitor and explosion to explain the observed diversity of SNe~Ia
is still in the very early stages \cite{hof98,umeda}. 
At this stage, high signal-to-noise observations of the
spectroscopic and photometric evolution of nearby SNe~Ia can provide
valuable information. Unfortunately, good photometric coverage around
maximum light and sufficient monitoring of the decline phase is
available only for very few nearby SNe~Ia, and thus it is very
important to enlarge the sample. SN~1996X is one such object.

\begin{figure}
\psfig{file=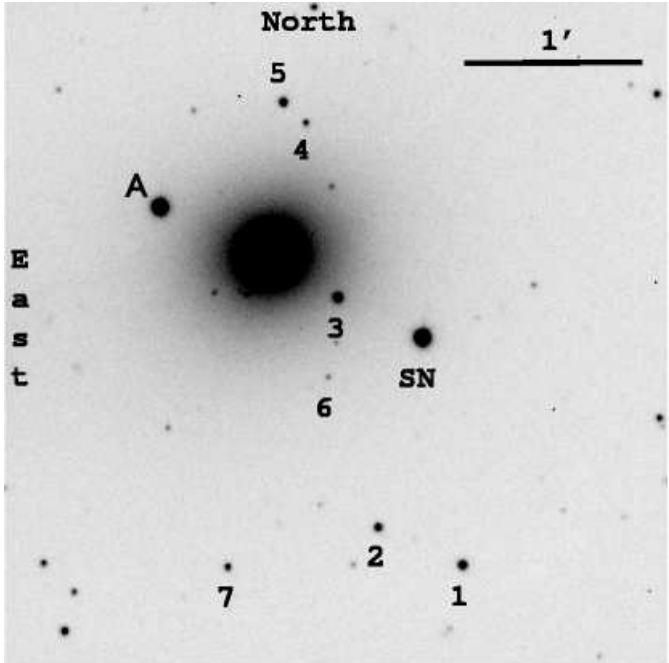,width=0.5\textwidth}
\caption{Identification chart for SN~1996X and for the stars of the local 
sequence (V band exposure obtained on Apr 15th, 1996 with the Dutch 0.9m 
telescope).}
\label{figsn}
\end{figure}

SN~1996X, wasdiscovered independently by Evans and Takamizawa on April 12.5 UT
(I.A.U. Circ. No. 6380). The SN was located at $R.A. = 13^{\rm h} 18 ^{\rm m}
01^{\rm s}.13$, DEC.$ = -26^{o}$ 50' 45".3 (equinox 2000.0), that is 52''W and
31''S of the nucleus of the elliptical galaxy NGC~5061 (Fig.~\ref{figsn}).
Early spectroscopic observations at different sites (cf. I.A.U. Circ. 6381)
showed that SN~1996X was a type Ia caught before maximum light, which 
motivated an extensive observational campaign with ESO telescopes.

SN~1996X was also observed at other observatories, in particular at CTIO
\cite{riess99} (\verb|http//www.ctio.noao.edu\~riccov/sn96x.gif|) and at
Mt. Whipple. A near-infrared spectrum was published by Bowers et al.\/
\shortcite{bowers}.

Interestingly, Wang, Wheeler \& H{\"o}flich \shortcite{wang} claimed
that SN~1996X showed evidence of intrinsic polarisation, suggesting an
asphericity of $\sim 10\%$ in the elements distribution in the region
of partial burning. This is the first time polarisation was measured
in a SN~Ia.

Finally, the SN field was observed at two late epochs with the WFPC2
on board HST (HST Data Archive).

In this paper we present the ESO observations, summarised in the UBVRIJ 
light curves and in the description of the evolution of the optical spectra. 
In addition we discuss some spectral modelling, and compare SN~1996X with 
other SNe~Ia in the framework of using SNe~Ia as distance indicators.

\section{Observations and data reduction}

\subsection{Photometry}

Photometry of SN~1996X was obtained mainly using the Dutch 0.9m telescope at
ESO - La Silla (CCD TK33; scale 0.44 arcsec/pixel). In addition we used the
ESO/MPI 2.2m telescope with EFOSC2 (CCDs TK19, 0.33 arcsec/pixel; Loral 40,
0.26 arcsec/pixel), the ESO 3.6m telescope with EFOSC1 (CCD TK26; 0.34
arcsec/pixel) and the ESO 1m DENIS telescope (CCD TK, 0.7 arcsec/pixel). Other
observations were obtained at Asiago with the 1.82m telescope (CCD TK512M; 0.34
arcsec/pixel). 

The frames were reduced using standard IRAF packages.  Photometric nights were
calibrated with observations of Landolt \shortcite{land} standard stars and
used to define a local standard sequence which, in turn, was used to calibrate
frames obtained under non-photometric conditions.  The stars of the local
sequence are identified in Fig.~\ref{figsn} and their magnitudes are listed in
Table~\ref{sequenza} along with r.m.s. deviations of the measurements obtained
on different nights.

Infrared photometry was obtained at the ESO/MPI 2.2m telescope, equipped with
IRAC2 (detector NICMOS-3 + lens LB, scale 0.278 arcsec/pixel), and at the ESO
1m DENIS telescope (NICMOS-3, 2.81 arcsec/pixel).

Calibration of the infrared photometry was performed as described in
Lidman, Gredel \& Moneti \shortcite{lidman}.  As a check for
photometric conditions we monitored star A (identified in Fig.~1),
which was measured at J $= 10.53\pm0.01$.

\begin{figure}
\psfig{file=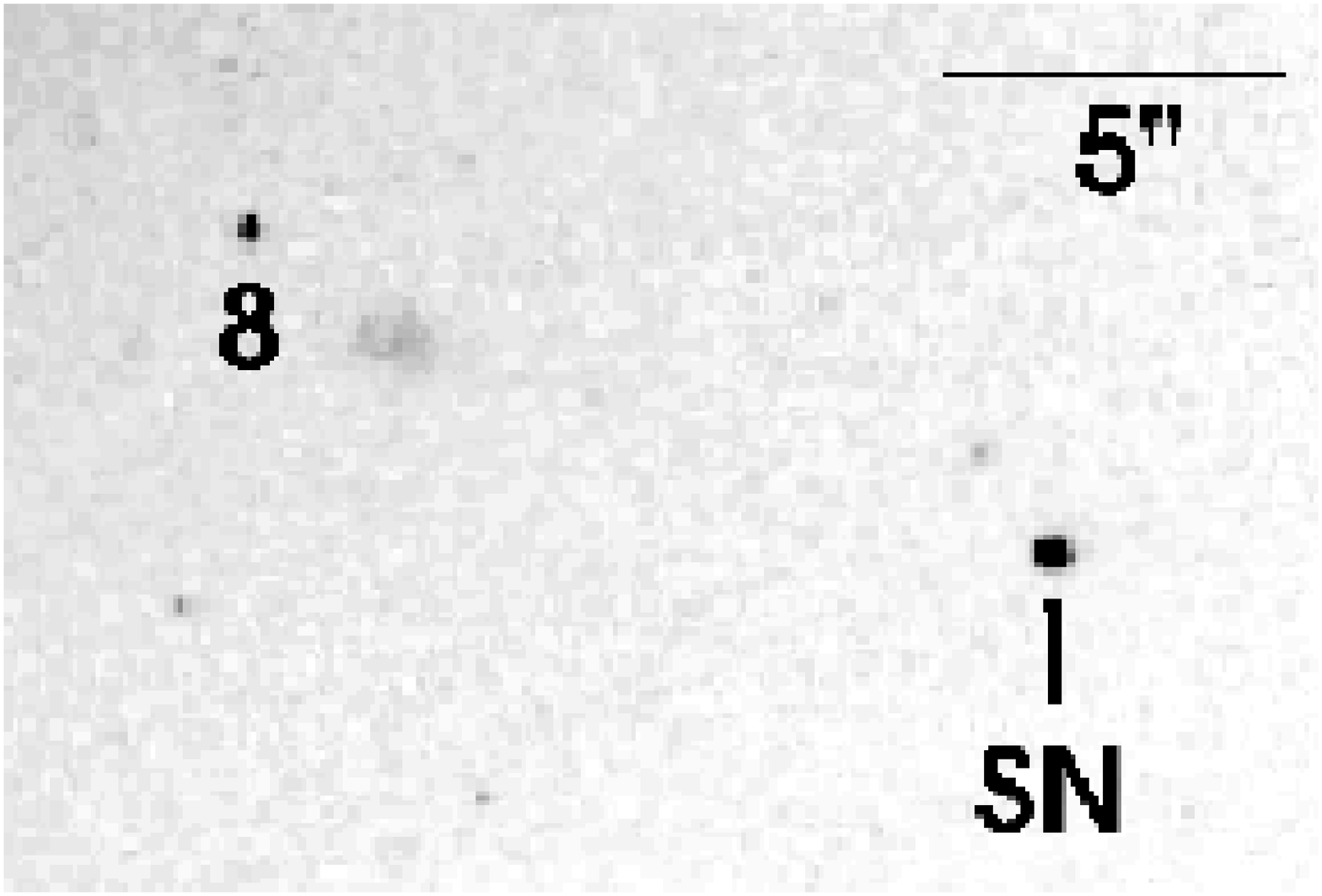,width=0.45\textwidth}
\psfig{file=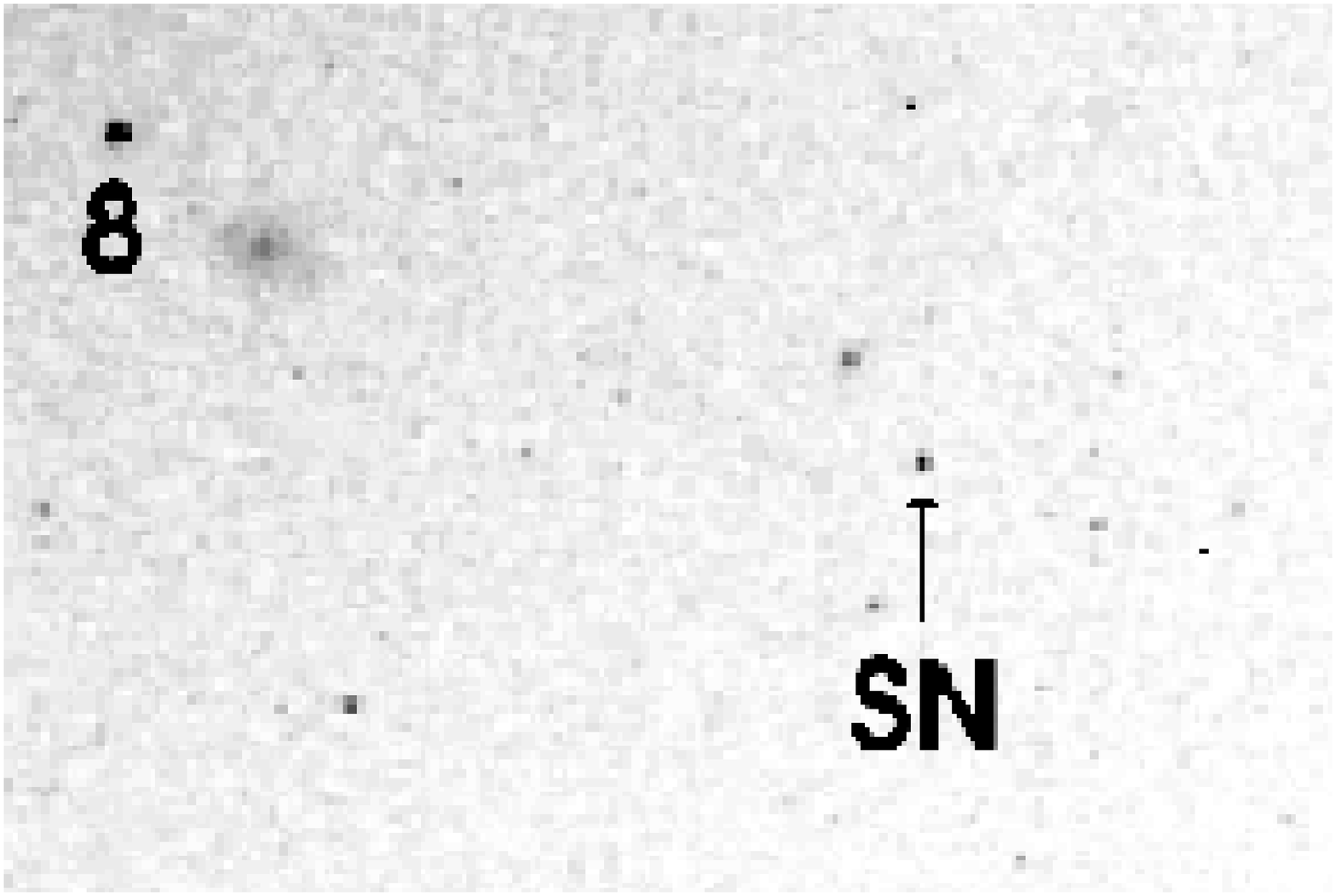,width=0.45\textwidth} 
\caption{WFPC2 observations of SN~1996X on 1997, Apr 5 (top panel)
and Sept 1 (bottom panel).}
\label{hst}
\end{figure}

\begin{table*}
\begin{center}
\caption{Magnitudes for the stars of the local sequence in the field 
of SN~1996X. The stars are identified in Fig.~1 and 2.}
\label{sequenza}
\begin{tabular}{cccccc}  
\hline
~~~id~~~  & U & B & V & R & I \\
\hline
{\bf 1} & 17.99$\pm$.02 & 17.26$\pm$.01 & 16.34$\pm$.01 & 15.78$\pm$.02 & 15.25$\pm$.01 \\
{\bf 2} & 18.35$\pm$.03 & 17.76$\pm$.01 & 16.90$\pm$.02 & 16.43$\pm$.01 & 15.97$\pm$.01 \\
{\bf 3} & 17.27$\pm$.02 & 16.82$\pm$.01 & 15.90$\pm$.01 & 15.38$\pm$.01 & 14.89$\pm$.01 \\
{\bf 4} & 18.89$\pm$.03 & 18.74$\pm$.01 & 17.94$\pm$.01 & 17.49$\pm$.01 & 17.08$\pm$.01 \\
{\bf 5} & 17.15$\pm$.02 & 17.16$\pm$.01 & 16.48$\pm$.02 & 16.04$\pm$.01 & 15.66$\pm$.03 \\
{\bf 6} & 19.52$\pm$.01 & 19.62$\pm$.02 & 19.02$\pm$.01 & 18.70$\pm$.02 & 18.26$\pm$.02 \\
{\bf 7} & 19.03$\pm$.03 & 18.51$\pm$.02 & 17.67$\pm$.01 & 17.16$\pm$.01 & 16.67$\pm$.01 \\
{\bf 8} &               & 23.44$\pm$.04 & 22.52$\pm$.04 &               & 21.24$\pm$.04   \\
\hline
\end{tabular}          
\end{center}
\end{table*}

The SN was located in a region of smoothly varying background, and thus 
reliable estimates of the SN magnitude could in principle be obtained with
plain aperture photometry. However, we preferred to use a point spread function
(PSF) fitting technique \cite{88h}, because: $i)$ the number of stars in the
field is sufficient to measure the PSF and $ii)$ this method is less sensitive
than aperture photometry to contamination by bad pixels and cosmic rays.

Observations of SN~1996X at late phases were obtained with the WFPC2 on board
HST on two epochs: April 5 (F555W and F814W filters) and September 1, 1997
(F439W, F450W, F555W and F606W filters).  The calibrated exposures were
retrieved from the ST-ECF HST archive, properly aligned and combined to
eliminate cosmic rays.  At the two epochs pointing and camera orientation were
quite different and the field overlap is only a small area around the SN.  
The instrumental magnitudes of the SN and of a foreground star (\#8 in the
Fig.\ref{hst}) were converted to the Landolt UBVRI system following the
prescription of Holtzman et al. \shortcite{holtzman}.  The Apr 5 observations
are bracketed by ground based observations, so that it was possible to verify
that, within the errors, the converted HST and ground-based photometry are
consistent.

The SN optical magnitude is reported in Table 2, and the IR ones in Table 3. 
The error estimates include the uncertainties in the internal calibration of
the local sequence and in the PSF fitting. The latter dominates at late phases
when the SN becomes faint.

\begin{table*}
\caption{UBVRI photometry for SN~1996X.}
\label{jour}      
\begin{tabular}{rrcccccl}       
\hline
Date & Phase$^*$ & U & B & V & R & I& Telescope  \\
     & [days] \\
\hline
14/04/96 & $-3.7$ & 12.71 $\pm$.03 & 13.29 $\pm$.05 & 13.30 $\pm$.02 & 13.3  $\pm$.02 & 13.42 $\pm$.03& ESO/Dutch 0.9m  \\
15/04/96 & $-2.7$ & 12.75 $\pm$.03 & 13.34 $\pm$.02 & 13.35 $\pm$.02 & 13.26 $\pm$.01 & 13.39 $\pm$.03& ESO/Dutch 0.9m  \\
16/04/96 & $-1.7$ & 12.79 $\pm$.02 & 13.27 $\pm$.01 & 13.29 $\pm$.02 & 13.25 $\pm$.01 & 13.38 $\pm$.01& ESO/Dutch 0.9m  \\
16/04/96 & $-1.6$ &   --           &   --           & 13.22 $\pm$.01 & 13.17 $\pm$.01 &            -- & ESO 3.6m+EFOSC  \\
17/04/96 & $-0.8$ & 12.77 $\pm$.04 & 13.25 $\pm$.01 & 13.28 $\pm$.01 & 13.23 $\pm$.02 & 13.43 $\pm$.01& ESO/Dutch 0.9m  \\
18/04/96 &  0.9  &   --           & 13.22 $\pm$.15 & 13.30 $\pm$.05 & 13.27 $\pm$.02 &   --          & Asiago 1.82m \\
19/04/96 & +1.9  & 12.94 $\pm$.05 & 13.30 $\pm$.01 & 13.21 $\pm$.01 & 13.18 $\pm$.01 & 13.54 $\pm$.02& ESO/Dutch 0.9m  \\
20/04/96 & +2.9  & 13.17 $\pm$.09 & 13.30 $\pm$.01 & 13.22 $\pm$.01 & 13.19 $\pm$.01 & 13.59 $\pm$.01& ESO/Dutch 0.9m  \\
23/04/96 & +5.9  &   --           & 13.51 $\pm$.01 & 13.41 $\pm$.01 & 13.34 $\pm$.01 & 13.81 $\pm$.02& ESO/Dutch 0.9m  \\
24/04/96 & +7.2  &   --           & 13.62 $\pm$.01 & 13.41 $\pm$.01 & 13.43 $\pm$.02 & 13.83 $\pm$.02& ESO/Dutch 0.9m  \\
25/04/96 & +8.2  &   --           & 13.70 $\pm$.01 & 13.45 $\pm$.01 & 13.51 $\pm$.01 & 13.97 $\pm$.01& ESO/Dutch 0.9m  \\
26/04/96 & +9.2  &   --           & 13.79 $\pm$.01 & 13.48 $\pm$.01 & 13.55 $\pm$.02 & 13.98 $\pm$.02& ESO/Dutch 0.9m  \\ 
27/04/96 & +10.1 &   --           & 13.91 $\pm$.01 & 13.58 $\pm$.01 & 13.60 $\pm$.01 & 14.00 $\pm$.01& ESO/Dutch 0.9m  \\ 
28/04/96 & +11.1 &   --           & 13.99 $\pm$.01 & 13.67 $\pm$.01 & 13.69 $\pm$.01 & 14.11 $\pm$.01& ESO/Dutch 0.9m  \\
04/05/96 & +15.9 &   --           &  --            &  --            &  --            & 14.11 $\pm$.02& ESO 1m + DENIS\\
05/05/96 & +16.9 &   --           &  --            &  --            &  --            & 14.11 $\pm$.02& ESO 1m + DENIS\\
05/05/96 & +17.0 & 15.15 $\pm$.02 & 14.83 $\pm$.03 & 14.11 $\pm$.02 & 14.07 $\pm$.02 & 14.16 $\pm$.02& ESO/Dutch 0.9m  \\
06/05/96 & +17.9 &   --           &  --            &  --            &  --            & 14.09 $\pm$.01& ESO 1m + DENIS\\ 
07/05/96 & +18.9 &   --           &  --            &  --            &  --            & 14.08 $\pm$.01& ESO 1m + DENIS\\
07/05/96 & +19.0 & 15.37 $\pm$.01 & 15.10 $\pm$.01 & 14.21 $\pm$.02 & 13.96 $\pm$.03 & 14.06 $\pm$.02& ESO/Dutch 0.9m  \\
09/05/96 & +22.0 & 15.73 $\pm$.02 & 15.29 $\pm$.01 & 14.23 $\pm$.01 & 14.00 $\pm$.02 & 13.88 $\pm$.01& ESO/Dutch 0.9m  \\ 
10/05/96 & +23.0 & 15.77 $\pm$.04 & 15.50 $\pm$.01 & 14.35 $\pm$.01 & 14.03 $\pm$.02 & 13.73 $\pm$.03& ESO/Dutch 0.9m  \\ 
11/05/96 & +24.0 &   --           &  --            &  --            &  --            & 14.00 $\pm$.01& ESO 1m + DENIS\\
12/05/96 & +25.0 &   --           &  --            &  --            &  --            & 13.97 $\pm$.01& ESO 1m + DENIS\\
13/05/96 & +26.3 &   --           &  --            &  --            & 14.14 $\pm$.02 & 13.87 $\pm$.03& ESO/Dutch 0.9m  \\
13/05/96 & +26.3 & 16.04 $\pm$.01 & 15.79 $\pm$.01 & 14.53 $\pm$.01 & 14.16 $\pm$.02 & 13.81 $\pm$.03& ESO/Dutch 0.9m  \\
14/05/96 & +27.2 & 16.20 $\pm$.01 &   --           & 14.68 $\pm$.01 & 14.22 $\pm$.01 & 13.97 $\pm$.02& ESO/Dutch 0.9m  \\
17/05/96 & +29.2 & 16.26 $\pm$.04 & 16.01 $\pm$.01 & 14.80 $\pm$.02 & 14.44 $\pm$.01 & 14.03 $\pm$.03& ESO/Dutch 0.9m  \\
19/05/96 & +31.3 & 16.25 $\pm$.16 & 16.09 $\pm$.03 & 15.01 $\pm$.02 & 14.51 $\pm$.01 & 14.13 $\pm$.02& ESO/MPI 2.2m + EFOSC2 \\ 
22/05/96 & +34.2 & 16.53 $\pm$.01 & 16.22 $\pm$.02 & 15.10 $\pm$.02 & 14.71 $\pm$.01 &   --          & ESO/Dutch 0.9m \\
26/05/96 & +39.2 & 16.75 $\pm$.02 & 16.44 $\pm$.01 & 15.36 $\pm$.01 & 14.95 $\pm$.01 & 14.64 $\pm$.01& ESO/Dutch 0.9m  \\ 
21/06/96 & +64.0 &   --           &    --          &   --           & 15.90 $\pm$.04 & 15.91 $\pm$.02& ESO/Dutch 0.9m  \\ 
24/06/96 & +67.1 &   --           &   --           & 16.12 $\pm$.01 & 16.01 $\pm$.01 & 16.10 $\pm$.01& ESO/Dutch 0.9m  \\
28/06/96 & +72.0 &   --           & 16.95 $\pm$.01 & 16.17 $\pm$.02 &   --           &     --        & ESO/Dutch 0.9m \\
15/07/96 & +89.2 &   --           & 17.08 $\pm$.02 & 16.67 $\pm$.01 & 16.61 $\pm$.01 & 16.81 $\pm$.01& ESO 3.6m + EFOSC\\
21/07/96 & +95.1 & 17.92 $\pm$.01 & 17.19 $\pm$.01 & 16.77 $\pm$.02 & 16.84 $\pm$.01 & 17.03 $\pm$.01& ESO/Dutch 0.9m  \\ 
06/09/96 & +142.1&   --           & 18.10 $\pm$.02 &   --           &   --           &   --          & ESO/MPI 2.2m + EFOSC2\\
02/01/97 & +259.3&   --           & 19.73 $\pm$.06 & 19.72 $\pm$.03 & 20.40 $\pm$.04 &  --           & ESO/Dutch 0.9m \\
10/02/97 & +298.5&   --           & 19.98 $\pm$.03 & 20.28 $\pm$.03 &   --           &   --          & ESO 3.6m + EFOSC\\
31/03/97 & +347.0&   --           & 21.00 $\pm$.09 & 21.06 $\pm$.07 & 21.63 $\pm$.12 &   --          & ESO/Dutch 0.9m \\
 5/04/97 & +351.6&   --           &  --            & 21.10 $\pm$.08 &  --            & 20.86 $\pm$.05& HST + WFPC2 \\
 5/04/97 & +351.6&   --           &  --            & 21.04 $\pm$.05 &  --            & 20.78 $\pm$.05& HST + WFPC2   \\
18/04/97 & +365.0&   --           & 21.36 $\pm$.10 & 21.34 $\pm$.09 &  --            &   --          & ESO/Dutch 0.9m \\
 1/09/97 & +501.2&   --           & 23.72 $\pm$.13 & 23.71 $\pm$.13 &  --            &   --     & HST + WFPC2\\
 1/09/97 & +501.2&   --           & 23.68 $\pm$.20 & 23.89 $\pm$.13 &  --            &   --     & HST + WFPC2\\
\hline
\end{tabular}

\flushleft $*$ Relative to the epoch of B maximum JD=2450191.5\\
\end{table*}

An external check on the reliability of the SN photometry can be obtained
by comparing our results with those recently published by Riess et al.\
\shortcite{riess99}. 

\begin{figure}
\psfig{file=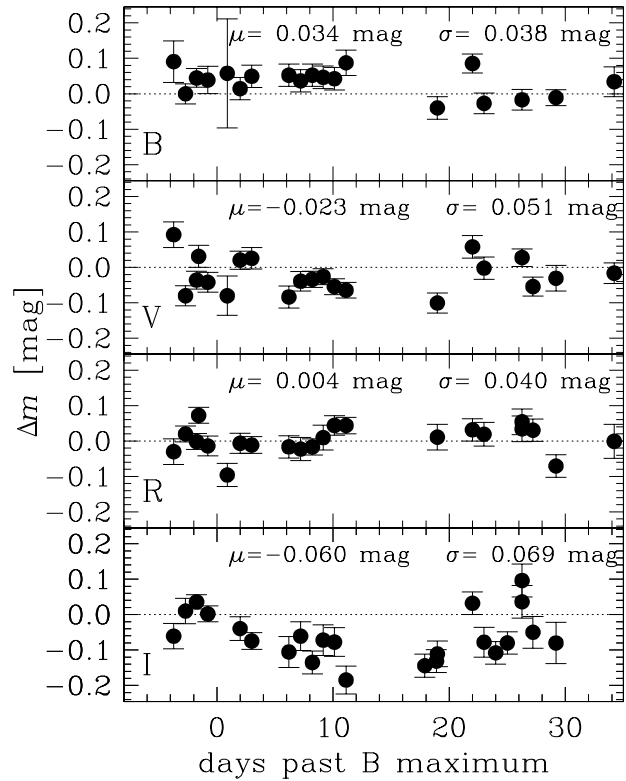,width=0.5\textwidth}
\caption{Comparison of the BVRI photometry of SN~1996X presented in this paper  
with that published by Riess et al.\ \shortcite{riess99}. Also indicated are
$\mu$ the average difference and $\sigma$ the r.m.s. around the mean.}
\label{cf_photo}                      
\end{figure}

This is done in Fig.~\ref{cf_photo}, where we plot for the various photometric
bands the difference between our estimates of the SN~1996X magnitudes and those
of Riess et al. These have been interpolated to the epochs of our observations
using cubic splines, but only if the phase difference was less than 3 days.
Error bars were obtained adding in quadrature the errors on the individual
points as reported for each source.

Although it is comforting that systematic differences are small (in all bands
$<\Delta m>\ \le 0.05$), the deviations of single points are larger than
expected from the combined errors. A similar conclusion was reached by Jha et
al.\ \shortcite{jha} when comparing their own photometry of SN~1998bu with 
that of Suntzeff et al. \shortcite{sunt99}. They attributed this problem to
differences in the filters used at different observatories (which are enhanced
by the peculiar and fast evolving SN spectra) and to the difficulty in
measuring the SN itself (see also Suntzeff 2000). Indeed, the deviations appear
more pronounced in the I band, which is difficult to define and maintain
properly because CCDs do not provide a fixed red-wavelength cut-off. The same
is true for the U band at the other end of the optical spectrum but, since
Riess et al. did not publish U photometry for SN~1996X, a similar test could
not be performed.

Therefore one should be wary of overinterpreting small differences in the 
light curves of SNe~Ia obtained with different instruments, especially for 
the U and I bands.

\begin{table} 
\begin{center}
\caption{Infrared photometry for SN 1996X.}
\label{ir}
\begin{tabular}{|c|c|c|c|c|}  
\hline
date & phase$^*$ & J & H & K' \\
     & [days] &  &   &    \\
\hline
\hline
 29/04/96 & +12.3 & 15.24$\pm$.05 & 14.19$\pm$.04 & 14.17$\pm$.06 \\ 
 04/05/96 & +16.0 & 15.05$\pm$.11 & -- & $\leq$ 14.0 \\ 
 05/05/96 & +16.9 & 15.02$\pm$.10 & -- & -- \\ 
 06/05/96 & +17.9 & 14.89$\pm$.10 & -- & -- \\ 
 07/05/96 & +18.9 & 15.12$\pm$.13 & -- & -- \\ 
 11/05/96 & +24.0 & 14.67$\pm$.07 & -- & -- \\ 
 12/05/96 & +25.0 & 14.56$\pm$.07 & -- & -- \\ 
 23/08/96 & +128.0 & -- & $\leq$ 17.0 & $\leq$ 15.0 \\
\hline
\end{tabular}
\end{center}

$*$ Relative to the epoch of B maximum JD=2450191.5\\
\end{table}

\subsection{Spectroscopy}

Most of the spectroscopic observations were obtained with the ESO~1.5m
telescope equipped with a Boller \& Chivens spectrograph.
% and the CCD Loral-Lesser \#39.  
Other observations were obtained using the
ESO/MPI 2.2m + EFOSC2 and the ESO 3.6m + EFOSC1. One spectrum was
obtained at Asiago with the 1.82m telescope and a B\&C spectrograph.

The journal of the spectroscopic observations is given in Table \ref{dati}.  
The overall spectral evolution is shown in Fig.~\ref{spec_seq}. 
Two of the spectra displayed are the result of merging spectra
obtained using different grisms during the same night.

Spectra were reduced using standard IRAF routines.  He-Ar lamp exposures were
used for wavelength calibration, and observations of spectrophotometric
standard stars \cite{hamuy_st1,hamuy_st2} for flux calibration. The absolute
flux calibration of the spectra was checked against the photometry and the
flux scale was adjusted if necessary.

\begin{table*}
\begin{center}
\caption{Journal of the spectroscopic observations.}
\label{dati}
\begin{tabular}{crccl} 
\hline
date & phase$^*$  & Range & Res. & Instrument\\
     &  [days]    & [\AA] & [\AA] & \\ 
\hline 
\hline 
14/04/96 & -3.8  & 4600-6700  & 2.5& ESO1.5+B\&C+gr10    \\
16/04/96 & -1.7  & 3700-7000  & 16 & ESO3.6+EF+B300,R300 \\
17/04/96 & -0.8  & 4600-6700  & 2.7& ESO1.5+B\&C+gr10   \\
18/04/96 &  0.2  & 3100-10700 & 18 & ESO1.5+B\&C+gr2    \\
19/04/96 &  1.2  & 3100-10700 & 15 & ESO1.5+B\&C+gr2    \\
21/04/96 & +3.9  & 4200-8600  & 30 & Asiago1.8+B\&C+150tr \\
25/04/96 & +7.2  & 3100-10500 & 11 & ESO1.5+B\&C+gr2    \\
30/04/96 & +12.1 & 4100-6800  & 14 & ESO1.5+B\&C+gr5    \\
10/05/96 & +22.0 & 3100-10600 & 11 & ESO1.5+B\&C+gr2    \\
12/05/96 & +24.3 & 3100-10600 & 11 & ESO1.5+B\&C+gr2    \\
19/05/96 & +31.3 & 3200-9100  & 11 & ESO2.2+EF2+gr6,5,1 \\
13/06/96 & +56.0 & 3100-10400 & 13 & ESO1.5+B\&C+gr2    \\
14/06/96 & +57.1 & 3100-10400 & 14 & ESO1.5+B\&C+gr2    \\
14/07/96 & +87.6 & 3100-10600 & 15 & ESO1.5+B\&C+gr2    \\
20/12/96 & +246.0& 3500-9300  & 8  & ESO1.5+B\&C+gr25   \\
10/02/97 & +298.4& 3700-6900  & 16 & ESO3.6+EF+B300     \\
\hline
\end{tabular}
\end{center}

$*$ Relative to the epoch of B maximum JD=2450191.5\\
\end{table*}

\section{Interstellar absorption}\label{abso}

SN~1996X exploded in the outskirts of an early-type galaxy, and therefore
extinction in the parent galaxy should be negligible.  Indeed, our higher
resolution spectra show no evidence of interstellar absorption lines at the
host galaxy rest frame.

However, a faint narrow NaI~D absorption (EW=0.27\AA) is detected at the rest 
wavelength. This can be attributed to interstellar gas in our own Galaxy. 
Indeed, Munari \& Switter \shortcite{munari}, studying
a sample of galactic early type stars, have shown that there is a
tight correlation between EW(NaI~D) and interstellar reddening.
Using their calibration we obtain for SN~1996X, $E(B-V) =
0.10\pm0.03$ mag, while a similar relation obtained by Barbon et al.
(1990) for SNe~Ia gives $E(B-V) = 0.07$ mag.
These values are in excellent agreement with other estimates of the
galactic absorption in the direction of NGC 5061.  Burstein \& Heiles
\shortcite{B&H} found ${\rm A}_{\rm B} = 0.25$ from H I/galaxy counts,
while Schlegel, Finkbeiner \& Davis \shortcite{schlegel} deduced ${\rm
A}_{\rm B}=0.30$ from the COBE/DIRBE and IRAS/ISSA maps. In this paper
we adopt ${\rm A}_{\rm B} = 0.30\pm 0.05$ as the fiducial value for
the total extinction.

\section[]{The light curves}

\begin{figure*}
\centerline{
\psfig{file=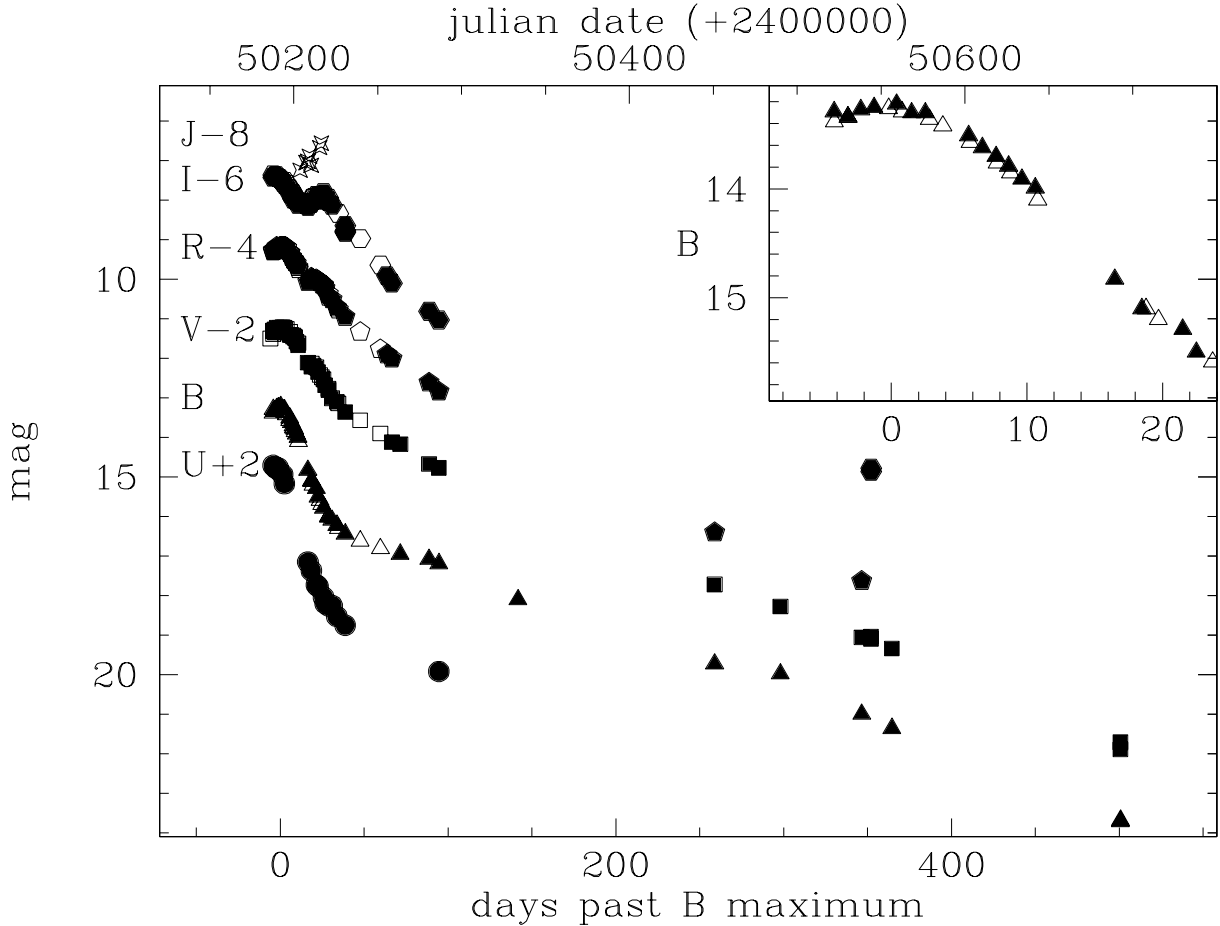,width=\textwidth,angle=270}
}
\caption{The UBVRIJ light curves of SN 1996X; ordinate scale refers to 
the B band, other bands are shifted by the amount reported in the legenda.  
Filled symbols are the values presented in this paper and open symbols are 
the estimates published by Riess et al.\ (1999).}
\label{lcur}                      
\end{figure*}              

The UBVRIJ light curves of SN~1996X are shown in Fig.~\ref{lcur}.  We 
estimated that the SN reached B maximum on Apr 18, 1996 (JD$=2450191.5\pm0.5$) 
at a magnitude $B=13.24\pm0.02$. Following the definition of Phillips
\shortcite{phil93}, we measured $\Delta m_{15}(B) = 1.31\pm.08$ mag for
SN~1996X, in good agreement with previous estimates, i.e. $\Delta
m_{15}(B)=1.28$ \cite{riess98a} and $\Delta m_{15}(B) = 1.25\pm0.05$
\cite{phil99}).
\footnote{the correction to $\Delta m_{15}(B)$ caused by the presence of
reddening (cf. sect.~\ref{redd}) is very small \cite{phil99} and has been 
neglected}.

The apparent magnitudes in the different bands at the epoch of the respective 
maxima are listed in Table~\ref{main}, along with other light curve
parameters.  As is normally observed in SN~Ia, V and R maxima occured 2-3d
later than B maximum, while the U and I maxima preceded it by 4d.

The I light curve shows the secondary maximum typical of ``normal'' SNe~Ia.
This occurs $~\sim25$ days after B maximum at a magnitude of 13.90, i.e. 0.50
mag fainter than the first maximum. At the same epochs a noticeable shoulder
appears in the R curve, and a weaker one in V. The few available J
observations appear to occur in the brightening phase to secondary maximum.

The light curves of SN~1996X are similar to those of other ``normal'' SNe~Ia,
e.g. SN~1994D or SN~1992A. The decline rate of SN~1996X is the same as that of
SN~1994D $\Delta m_{15}(B) = 1.32 \pm 0.05$, and slightly smaller than for
SN~1992A, $\Delta m_{15}(B) = 1.47 \pm 0.05$ mag \cite{phil99}.

\begin{figure}        
\psfig{file=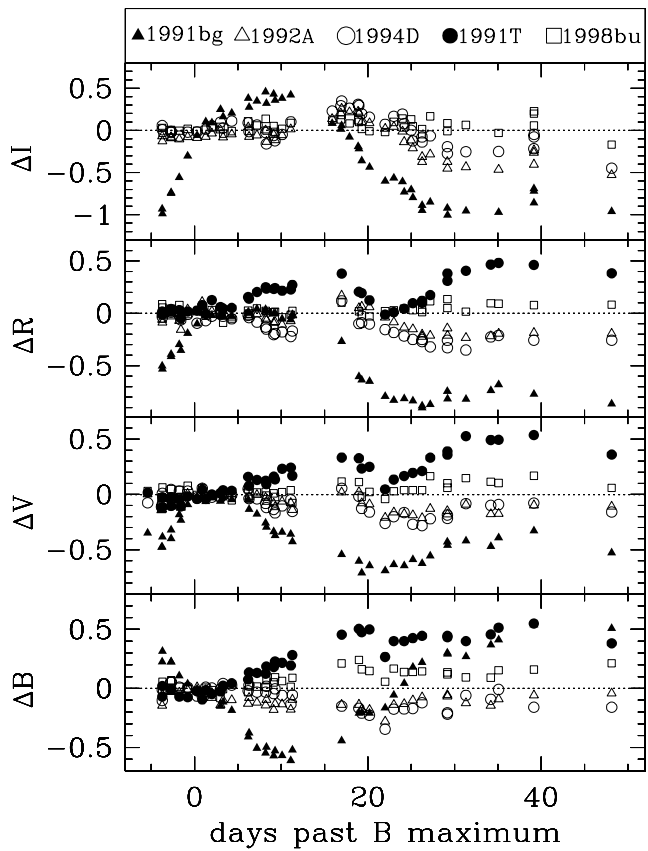,width=0.5\textwidth}
\caption{Difference of the light curve of SN~1996X with that of 
representative SN~Ia. For the comparison the light curves were normalised
to the magnitude and epoch of maximum.}
\label{comp}               
\end{figure}

In Fig.~\ref{comp} we plot the difference between the photometry of SN~1996X
and those of other well studied SNe~Ia. The light curves were normalised to the
apparent maximum in the given band and to the epoch of B maximum, and the
differences were computed relative to the phase of SN~1996X observations (the
light curves of the other SNe were interpolated when necessary).  We selected
as comparison the normal SNe 1992A and 1994D, and also SNe 1991bg ($\Delta
m_{15}(B) = 1.93\pm0.10$), 1998bu ($\Delta m_{15}(B) = 1.01\pm0.05$) and
1991T ($\Delta m_{15}(B) = 0.94\pm0.05$) to cover the entire range of SN~Ia 
luminosity evolution.

Some systematic differences are apparent from this figure.  In the B band the
difference is greatest at around phase 20d, and then it remains constant or
decreases. This is of course the reason why $\Delta m_{15}(B)$ is so effective
in discriminating between SNe~Ia light curves.  In the V and R bands, the
picture is less clear.  Whereas in the B band the SNe closest to SN~1996X  are
SNe 1992A and 1994D, as expected because of the similar $\Delta m_{15}(B)$, in
the R band the best match to SN~1996X is SN~1998bu.  This is true also in the 
I band. Indeed in SNe 1996X and 1998bu the I secondary maximum occurred at the
same phase, i.e. 4-5 days later than in SN~1994D and was 0.15 mag fainter with
respect to the first maximum.
 
SN 1996X is one of the very few SNe for which observations at epochs
larger than 1 yr are available.  The importance of the late light
curve is that it provides useful insights into the process of
radioactive energy deposition.  The photometry of SN~1996X at day 500
is particularly accurate because it was obtained with HST, whose
superior angular resolution allows a positive target identification
and a more accurate background subtraction. It is also important that
observations were obtained in different bands, giving an estimate of
the B-V colour. The result is ${\rm B}-{\rm V}_{(500d)}=-0.21\pm0.24$,
which shows that the colour, and thus possibly the spectral shape as
well, did not change significantly after day 150.

After day 150, and up to the epoch of the last observation, the light curves
are remarkably linear in all bands. The decline rate is 1.7 mag (100d)$^{-1}$
in B and V, and slightly smaller (1.4 mag (100d)$^{-1}$) in R, although the
photometric coverage in this band is not so good.

\begin{figure}
\centerline{
\psfig{file=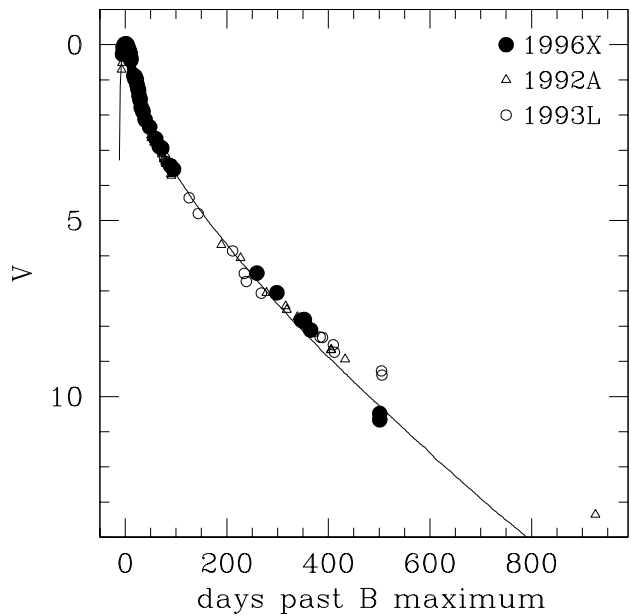,width=0.5\textwidth}
}
\caption{The V late light curve of SN~1996X is compared with that of
SN 1993L and 1992A (cf. Cappellaro et al. 1997) and with a W7 light
curve model (cf.\ref{bol_sec}) }
\label{latelc}                      
\end{figure}

A comparison between the late V light curves of SN~1996X and those of the
SNe~Ia 1993L and 1992A \cite{cap97}, normalised to the epoch and magnitude of
maximum, is shown in Fig.~\ref{latelc}.  Up to day 400 the three light curves
are similar, but later on SN~1996X declines at a much faster rate than the
other two objects.  The luminosity decline of SNe 1992A and 1993L is very close
to the decline of energy release from $^{56}$Co decay, which translates to 0.98
mag (100d)$^{-1}$. Since at such late phases $\gamma$-ray deposition is very
small, this pattern may be taken to indicate that complete deposition of
positrons is the principal source of power for the light curve. Indeed,
Cappellaro et al. \shortcite{cap97} could fit one very late point (1000d) 
in mthe light curve  of SN~1992A using a very large positron opacity, thus
suggesting complete positron deposition. On the other hand, a normal value
$\kappa_{e^+} = 7$cm$^2$ g$^{-1}$ can successfully explain the point at 500d 
in the light curve of SN~1996X.

\begin{table*}
\caption{SN 1996X photometric parameters: the apparent magnitudes and epochs 
of maximum in the different bands; the colours at the epoch of B maximum; 
the decline rate $\Delta m_{15}^B$ and $\Delta m_{20}^V$ (Hamuy et al. 1996), 
and the average decline rate (units are mag (100d)$^{-1}$) in three different 
section of the light curve, namely 5d-30d ($\beta$), 50d-150d ($\gamma_1$) 
and $>$150d ($\gamma_2$).}
\label{main}
\begin{tabular}{|c|c|c|c|c|c|} 
\hline
maximum       &    U          &   B           &     V         &   R           &  I           \\
JD (+2450000) & $188.0\pm1.0$ & $191.5\pm0.5$ & $193.5\pm0.5$ & $194.5\pm0.5$ & $188.0\pm1.0$\\
$mag$         & $12.73\pm.03$ & $13.24\pm.02$ & $13.21\pm.01$ & $13.19\pm.01$ & $13.38\pm.01$\\
\\
$\Delta m_{15}$ &             & $1.31\pm.05$ \\
$\Delta m_{20}$ &             &              & $1.05\pm.05$  & \\
\\
$\beta$        & 12.0$\pm$.6  &  $11.4\pm.3$  &  6.2$\pm$.1 & 4.0$\pm$.3 & 0.21$\pm.2$ \\
$\gamma_1$     &  2.1$\pm$.2  &  $1.60\pm.10$ & 2.52$\pm$.09& 2.96$\pm$.07 &3.7$\pm.2$\\         
$\gamma_2$     &     --       &  $1.71\pm.05$ & 1.73$\pm$.05& 1.40$\pm$.2  &   --       \\
\\
colours$^\#$   &   U$-$B        &   B$-$V       &     V$-$R    &   V$-$I     &   \\
               & $-0.45\pm.05$  & $0.00\pm.03$  & $0.04\pm.02$ & $-0.20\pm.05$  &     \\
\hline
\end{tabular}

\# Measured at the epoch of B maximum
\end{table*}            

\section{The colour curves}\label{redd}

With the remarkable exception of the few ``subluminous'' SNe~Ia,
e.g. 1991bg and 1997cn (Turatto et al. 1996,1998), the colour curves
of normal SNe Ia have similar shapes.  Indeed, the $B-V$ colour evolution
of SN~1996X is very similar to those of the ``normal'' SNe 1992A and
1994D but also to that of the bright SN~1991T, whose luminosity
decline was much slower ($\Delta m_{15}B = 0.94$ mag, Phillips 1993).

However, a detailed examination of the early phases shows that even after 
correction for galactic reddening small shifts remain between the colour 
curves of different SNe. This may be attributed to reddening in the host 
galaxy or to intrinsic differences between the SNe or more likely to both 
effects. However, the discrepancy between the photometry from different groups
shown in Fig~\ref{cf_photo} and in Jha et al. \shortcite{jha} suggests that
differences in colour smaller than 0.1 mag should not be overinterpreted. 
This implies that the comparison of the colour curves cannot yield a reliable
estimate of the absorption if this is smaller than A$_B \sim 0.5$ mag.  
For SN~1996X all we can say is that the colour curve is consistent with 
negligible reddening in the SN host galaxy.

\begin{figure*}                 
\centerline{
\psfig{file=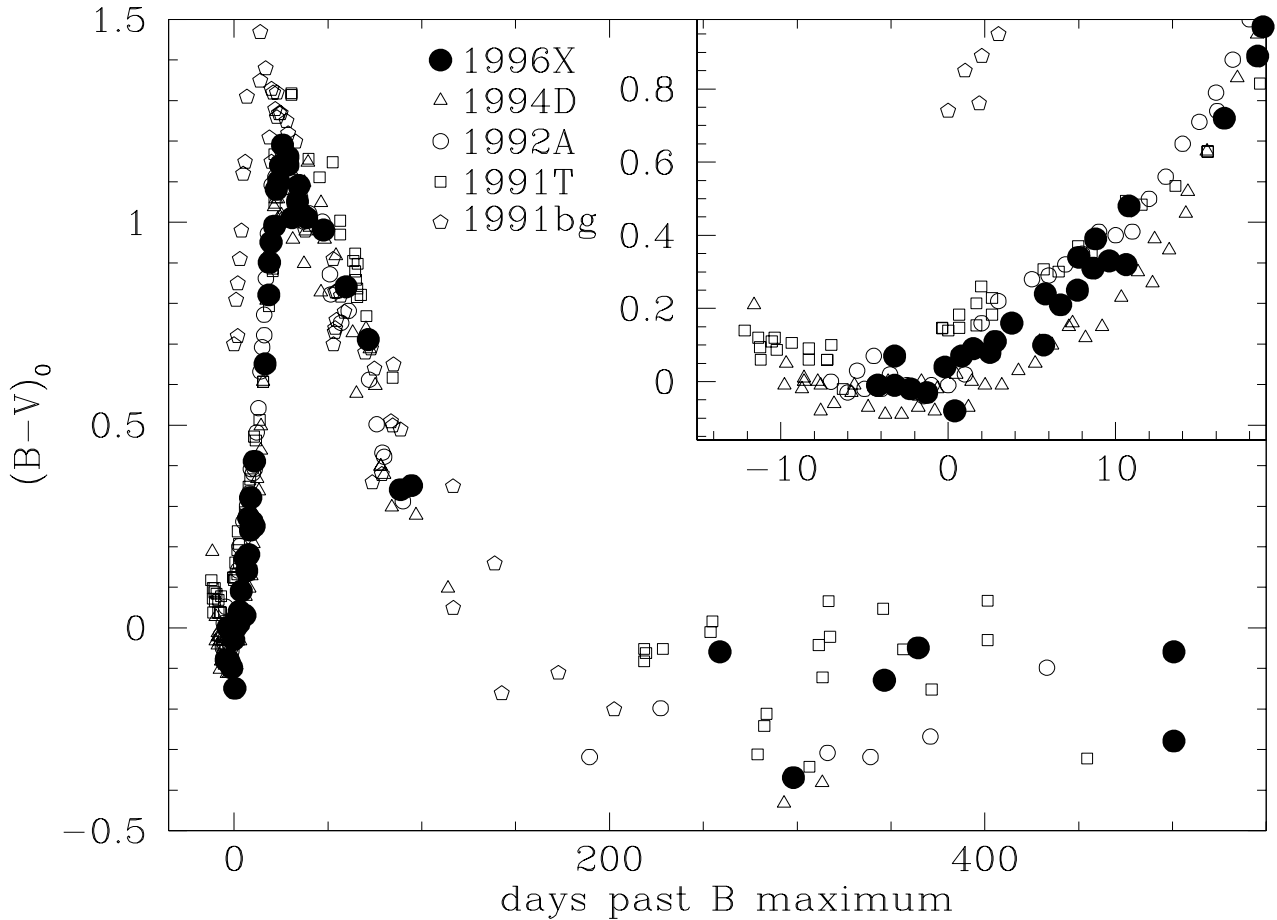,width=\textwidth,angle=270}
}
\caption{A comparison of the B-V colour curves of SN 1996X and other SNe Ia. In 
all cases only the galactic reddening (Schlegel et al. 1998) has been removed.}
\label{ccur}
\end{figure*}

\section[]{The absolute magnitude of SN~1996X}

The absolute magnitude of SN 1996X can be derived from the distance of the
parent galaxy or conversely we can use the relation between light curve shape
and SN absolute magnitude and hence derive the distance.

An estimate of the distance to NGC 5061 can be obtained from the recession
velocity and the Hubble law. After correcting the measured heliocentric
velocity (2065$\pm$46 km s$^{-1}$, \cite{dacosta}) for the Local Group 
infall onto Virgo (1949 km s$^{-1}$) and using ${\rm H}_0 = 65
{\rm km\,s}^{-1} {\rm Mpc}^{-1}$ we obtain a distance modulus $\mu = 32.38$.

Faber et al. \shortcite{faber} estimated the distance using the $D_n - \sigma$
relation. On the same ${\rm H}_0 = 65 {\rm km~s}^{-1} {\rm Mpc}^{-1}$ distance
scale they obtained a much smaller value, $\mu = 31.32\pm0.4$.

\begin{table}
\caption{Estimates of the distance of NGC 5061 and the magnitude of SN 1996X 
according to various authors.}
\begin{tabular}{lcc}   
\hline
Source & $\mu$ & M(B) \\
\hline
$D_n$ - $\sigma$ Faber  et al. (1989) & $31.32\pm0.4$    & $-18.38\pm0.40$ \\
   v$_{\rm Local~Group}$             & $32.80\pm0.4$    & $-19.86\pm0.40$ \\
$\Delta m_{15}$(B) \& Hamuy et al. (1996) & $32.02\pm0.2$  & $-19.08\pm0.21$ \\
$\Delta m_{15}$(B) \& Phillips et al. (1999) & $32.42\pm0.2$    & $-19.48\pm0.21$ \\
Riess et al. (1998a,1998b)   & 32.4             & $-19.16$ \\
\hline
\end{tabular}  

\label{dist}
\end{table}            

Alternatively, we can use the relation between $M_B$ vs $\Delta m_{15}$ to
calibrate the SN absolute magnitude.  Depending on the calibration adopted, we
obtain M$_B = -19.08\pm0.21$, \cite{hamuy_absmag} or M$_B = -19.48\pm0.21$
\cite{phil99} which corresponds to a distance modulus of $\mu = 32.02\pm0.2$
and $\mu = 32.42\pm0.2$ respectively. The large offset between the two
calibrations has been attributed to the use of different recipes for the
correction of both galactic and host galaxy extinction \cite{phil99}.

The absolute magnitude of SNe~Ia can also be calibrated using the {\em
Multicolour Light-Curve Shapes} (MLCS) method (Riess et al. 1996, 1998a). In
this approach, the light curve of the SN under study is compared with those 
of template objects to obtain a simultaneous estimate of the ``luminosity
correction'' $\Delta$ and of the extinction.  Although from a statistical 
point of view this method is expected to be more robust than the simple 
$M_B - \Delta m_{15}$ relation, the results are actually very sensitive to 
the calibration of the template objects, in particular the adopted distance 
and reddening.  This is because the purported monotonic relation between 
absolute magnitude and colour, with intrinsically brighter SN being bluer, 
is not yet fully confirmed.  This is manifested in the large difference
in the template colour curve calibration between Riess et al. 
\shortcite{riess} and Riess at al. \shortcite{riess98a}.

With reference to SN~1996X, Riess et al. (1998a; their Fig.~13) showed that the
MLCS fit of the V and B-V curves yields $\Delta = +0.25$, $E(B-V) = 0.0$ and a
calibrated absolute magnitude $M_B = -19.15$. Therefore the distance to
NGC~5061 is derived to be $\mu = 32.4$ (see also Riess et al. 1998b).  This is
the same result as obtained from the latest calibration of the $M_B - \Delta
m_{15}$ relation \cite{phil99}.  However there is an inconsistency with regard
to the total  observed reddening, which was estimated to be negligible whereas
the evidence for significant galactic reddening is compelling.
(cf.\ref{abso}).  Some inconsistencies in the determination of the reddening
using the MLCS method were also noticed by Turatto et al. (1998) in the 
case of SNe 1997cn and 1991bg.

The implication of the difference of the SN absolute magnitude
calibration for the modeling will be addressed in the next sections. 

Given that NGC 5061 is an elliptical galaxy, a further independent
determination of the distance could be obtained using the globular cluster
luminosity function method.

\section{The bolometric light curve}\label{bol_sec}

\begin{figure}        
\centerline{
\psfig{file=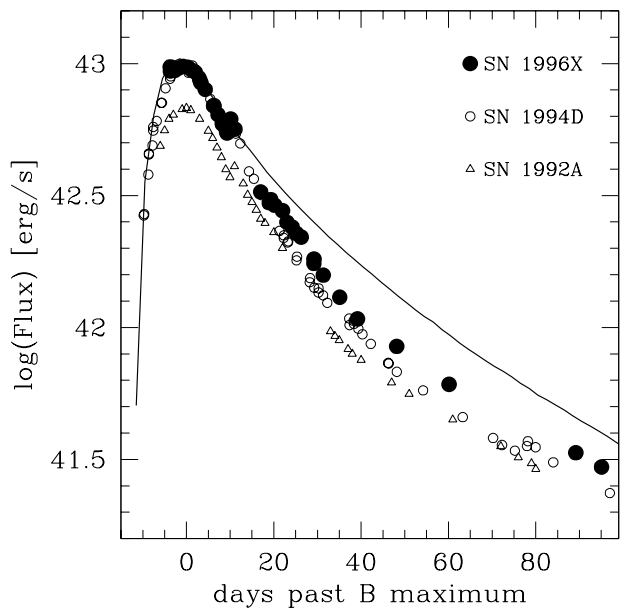,width=.50\textwidth}
}
\caption{The ``UVOIR'' bolometric light curve of SN 1996X ($\mu=32.02$, 
A$_{\rm B}$ = 0.30) compared with that of SN 1994D ($\mu=30.69$,
A$_{\rm B}$ = 0.24), SN~1992A ($\mu=31.35$, A$_{\rm B}$ = 0.0) and with
a W7 light curve model (cf. sec \ref{bol_sec}) }
\label{bol}                      
\end{figure}              

Combining HST UV data, UBVIR data in the optical and JHK data in the IR for
SN~1992A, Suntzeff \shortcite{sunt_92a} computed the so-called {\em uvoir}
bolometric luminosity, which integrates the flux emitted in the range 200-2200
nm, but only up to day 77.  In all other cases the bolometric light curve of
SN~Ia was calculated from data with incomplete wavelength coverage assuming
that ``normal SNe~Ia'' behave similarly.  This is the approach we adopt here to
derive the {\em uvoir} bolometric light curve of SN~1996X, since only
optical-band observations are available. We integrated the flux emitted in the
Johnson-Cousins U,B,V,R,I passbands at the epochs for which measurements are
available (for a few epochs the U magnitude was interpolated between points
adjacent in time).  This was then corrected to the 200-2200 nm adopting the
appropriate ``bolometric'' correction as given in Suntzeff \shortcite{sunt_92a}
for SN~1992A.

As a justification for using this approach we note that the optical
spectra and the light curve of SN~1996X are similar to those of
SN~1992A. It is also important that at any epoch the UV and infrared
emission account for at most 25$\%$ of the total flux.

For at least the one epoch when J, H and K' photometry for SN~1996X is
available we could measure the IR luminosity contribution.  We find that on
day +12 the infrared luminosity is about 10$\%$ of the total {\em uvoir}
luminosity, in very good agreement with the estimates of Suntzeff
\shortcite{sunt_92a} for the same epoch. Day 77 is the latest epoch for which
Suntzeff gives a ``bolometric correction'' for the optical flux. For later
epochs a constant optical to bolometric correction was adopted.

The {\em uvoir} bolometric light curve for the first 100d, corrected for
galactic reddening and for a distance modulus $\mu = 32.02$ is plotted in
Fig.\ref{bol} (if we adopted $\mu = 32.42$ the bolometric light curve would
peak at $\log L = 43.15$).  For comparison we also show the {\em uvoir}
bolometric light curve of SN 1994D obtained using the data of Patat et al
(1996) and assuming $\mu = 30.68$ and $A_B= 0.24$.

The observed light curves of SN~1996X were compared to a synthetic light curve
obtained with the simple Monte Carlo light curve code used in Cappellaro et al.
(1997). The synthetic bolometric light curve was computed for a W7 density
structure, with an ejecta mass of $1.38 M_\odot$ and a synthesised $^{56}$Ni
mass of $0.6 M_\odot$. The synthetic light curve reproduces the {\em uvoir}
bolometric light curves of both SNe 1994D and 1996X quite well near maximum. 
On the other hand, if we adopt the Phillips et al. \shortcite{phil99}
calibration (M$_B = -19.48$), the fit of the light curve peak requires a Ni
mass of $0.85 M_\odot$. This estimate, does not  depend much on the particular
model of the ejecta, would place SN~1996X near the high-Ni mass end of
Chandrasekhar mass models of SNe~Ia. If we consider that for its decline rate
SN~1996X is expected to be fainter than the average of `normal' SNe Ia, other
SNe would be very difficult to explain within the Chandrasekhar mass explosion
scenario.

About two weeks past maximum the observed luminosity decline becomes 
steeper. This is an epoch when the rapidly decreasing temperature may
lead to a decreasing optical opacity. Also, the opacity is probably a
function of element abundance in the ejecta, and the combination of
these two effects is possibly at the basis of the observed ``Phillips
relation'' between $M_B(Max)$ and $\Delta m_B(15)$. Since the optical
opacity was treated as constant in our code
($\kappa_{opt}=0.15$cm$^2$g$^{-1}$), this behaviour is not reproduced.

At later epochs ($t \geq 300$d) positron deposition becomes the dominant source
of power for the light curve. The synthetic curve was computed using the
standard value for the positron opacity, $\kappa_{e^+} = 7$cm$^2$ g$^{-1}$.  
As the bolometric correction at these phases is unknown, we compared the 
model with the V observations, after matching the magnitude at maximum
(Fig.~\ref{latelc}). This comparison is justified if the bolometric correction
is constant which, for epochs $>150$ days, is supported by the fact that the
colour remains costant.  It results that the model reproduces the V light curve
of SN~1996X between 250 and 500 days rather well, considering also the large
errors in the photometry at these advanced epochs. Both SNe 1992A and 1993L lie
well above the synthetic light curve at $t > 400$d.  As already discussed by
Cappellaro et al. (1997), this difference may imply that positron deposition
was more efficient in those SNe than in SN~1996X.  Whether this requires
different explosion models, magnetic field characteristics or degree of
ionization is not yet clear \cite{milne}.

\section{The spectral evolution}

Spectra of SN~1996X have been obtained at sixteen epochs, from phase --4d to
+298d, following the rapid evolution during the early phases in detail and
sampling more sparsely the slow late evolution. They are presented in Fig.~9.

The spectral evolution of SN~1996X is similar to that of other normal SNe~Ia.  
At early phases the continuum is very blue and it is dominated by lines
attributed to Fe-group (Fe, Co) and intermediate mass elements (in particular 
Si, Ca, S, Mg). The lines exhibit the characteristic P-Cygni profiles with the
minima of the absorption components shifting towards longer wavelengths
(smaller expansion velocities) with time. At two-three months after maximum the
continuum is significantly redder ($T_{BB} \sim$ 5-6000$^\circ$K), and nebular
lines of [FeII] and [FeIII] begin to appear. At about 300 days these are the 
dominant features in the spectrum.

Exceptionally bright or dim SNe~Ia (eg. SN~1991T and SN~1991bg) can easily be 
distinguished on the basis of their spectra, but differences also exist among
`normal' SNe~Ia. These are more evident before maximum and in the
nebular phase, but they disappear in the 2-3 months following maximum
(cf. Li et al., 1999). In any case, differences among the spectra of `normal'
SNe~Ia are small and can be appreciated only when observations of high
signal to noise are compared.
                 
In Fig. \ref{spc} we compare the spectra of SN~1996X with those of
two SNe with a similar luminosity decline, SN~1992A and SN~1994D. The
comparison is made at three different epochs: near maximum, two months
and 300 days after maximum. The spectra are {\em almost}
identical. One difference is that near maximum the 6355 \AA\/ SiII
absorption in SN~1992A is more blue-shifted than in both SN~1996X and
SN~1994D. This indicates that the average SiII velocity is larger by
$\sim 1200$ km s$^{-1}$ in SN~1992A. A similar difference in velocity is seen
in other SiII lines, e.g. the 5958-5979 \AA\ line, which produces the
weak absorption near 5800\AA, but not in lines of SiIII or SII.

Another difference near maximum is a faint line at 4950\AA, which appears to be
stronger, or better resolved, in the spectrum of SN~1996X. This feature is
probably due mostly to FeIII lines, and may indicate a higher degree of
ionization in SN~1996X compared to the other two SNe.

Finally, the ratio of the deep FeIII absorption at 4250\AA\ and of the
neighbouring weak SiIII line near 4400\AA\ is very different in the
three SNe.  This may be a temperature effect, since the SiIII line
requires a much higher ionisation/excitation temperature, or an
abundance effect, deserving further quantitative investigation.

\begin{figure*} 
\begin{center}
\psfig{file=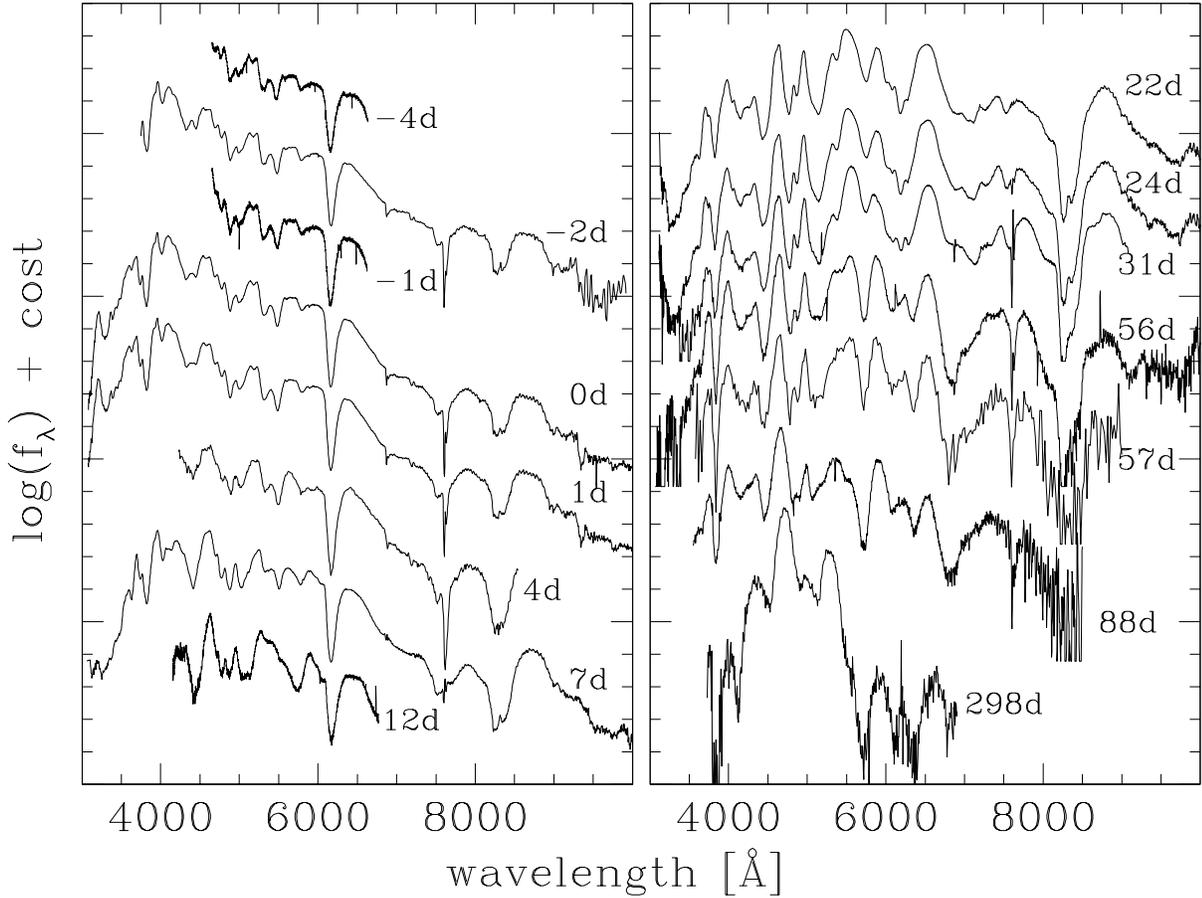,width=\textwidth,angle=270}
\caption{The spectral evolution of SN 1996X. Wavelength scale is in the 
observer rest frame.}\label{spec_seq}
\end{center}                                             
\end{figure*}    

The line velocities of Si II (6355 \AA) and Ca II H\&K have been measured from
the minima of the P-Cygni absorption features, after correcting the spectra for
the heliocentric velocity of the galaxy. These values can be taken to represent
approximately the photospheric velocity.  The velocities are listed in Table
\ref{vel} and plotted in Fig. 11.  They are similar to those of other normal
type SNe~Ia at equivalent phases (Barbon et al. 1990, Turatto et al. 1996). 
Closer inspection (Fig.~\ref{spc}) shows that whereas near maximum the
expansion velocity of SN~1996X is similar to that of SN~1994D, and 10$\%$
smaller than that of SN~1992A, the velocity decrease of SN~1996X is slow, so
that two weeks later the velocities of SN~1996X are comparable to those of
SN~1992A.

Lentz et al. (1999) computed emergent photospheric-phase spectra for a
grid of SN~Ia atmospheres, and argued that a blueshift of the SiII
6355\AA\ line could be indicative of an increased metallicity of the
SN~Ia progenitor. To reproduce the observed range in expansion
velocity with the Lentz et al. (1999) models, a change of the
metallicity by 2 order of magnitudes, from 0.1 to 10 times solar, is
required.  It should be noted that the parent galaxies of the three
SNe shown in Fig.~\ref{vl3} are all of early type, either S0 (for SNe
1992A and 1994D) or E (for 1996X), and variations by two orders of 
magnitude in metallicity between those galaxies are not expected.

\begin{figure} 
\begin{center}
\psfig{file=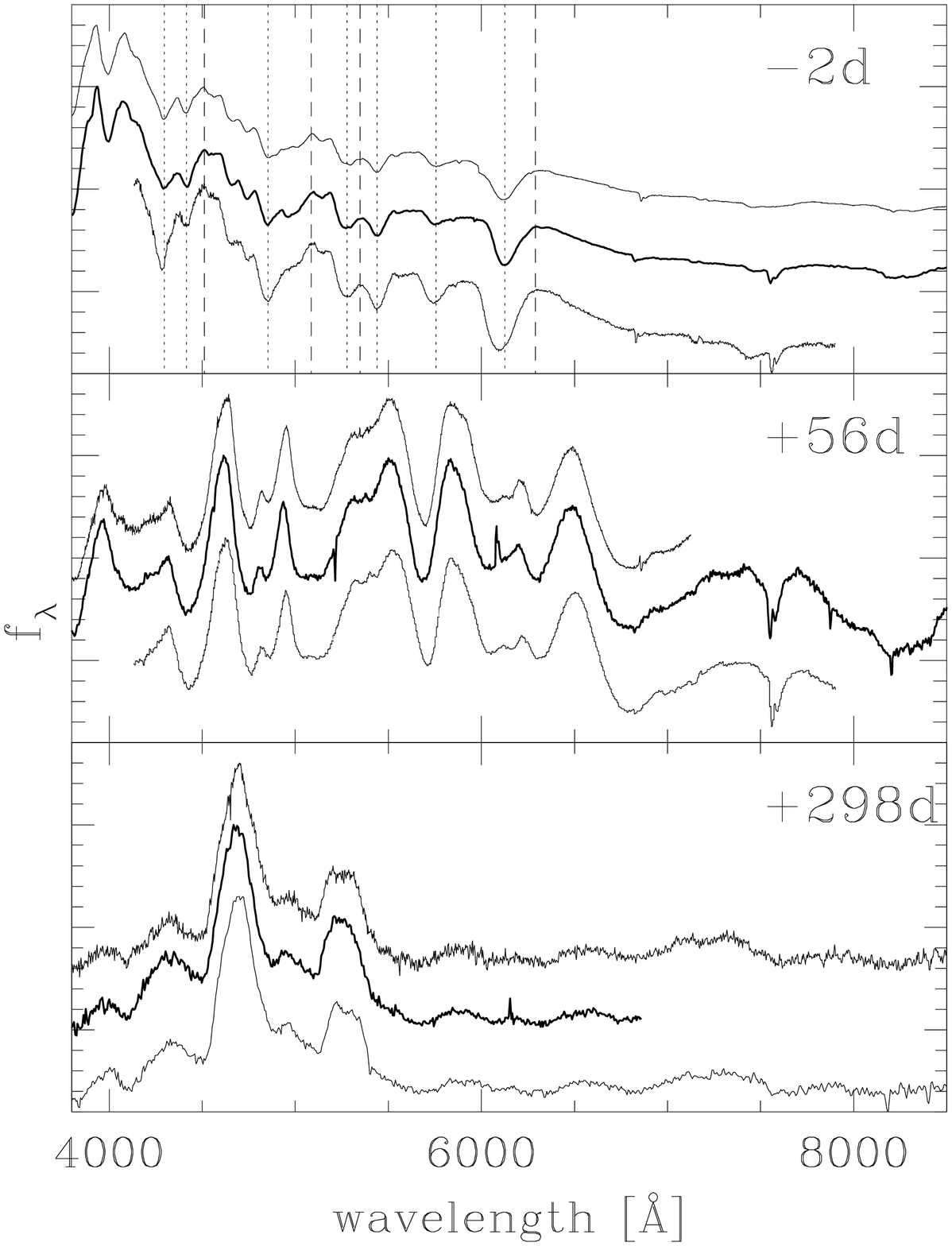,width=0.5\textwidth}
\caption{Comparison of spectra of SN 1996X (middle) at different 
phases with that of SN 1994D (top) and SN 1992A (bottom) at similar
phases.  Vertical lines in the top panel trace the position of the main
absorption and emission features.}\label{spc}
\end{center}
\end{figure}

\begin{figure} 
\begin{center}   
\psfig{file=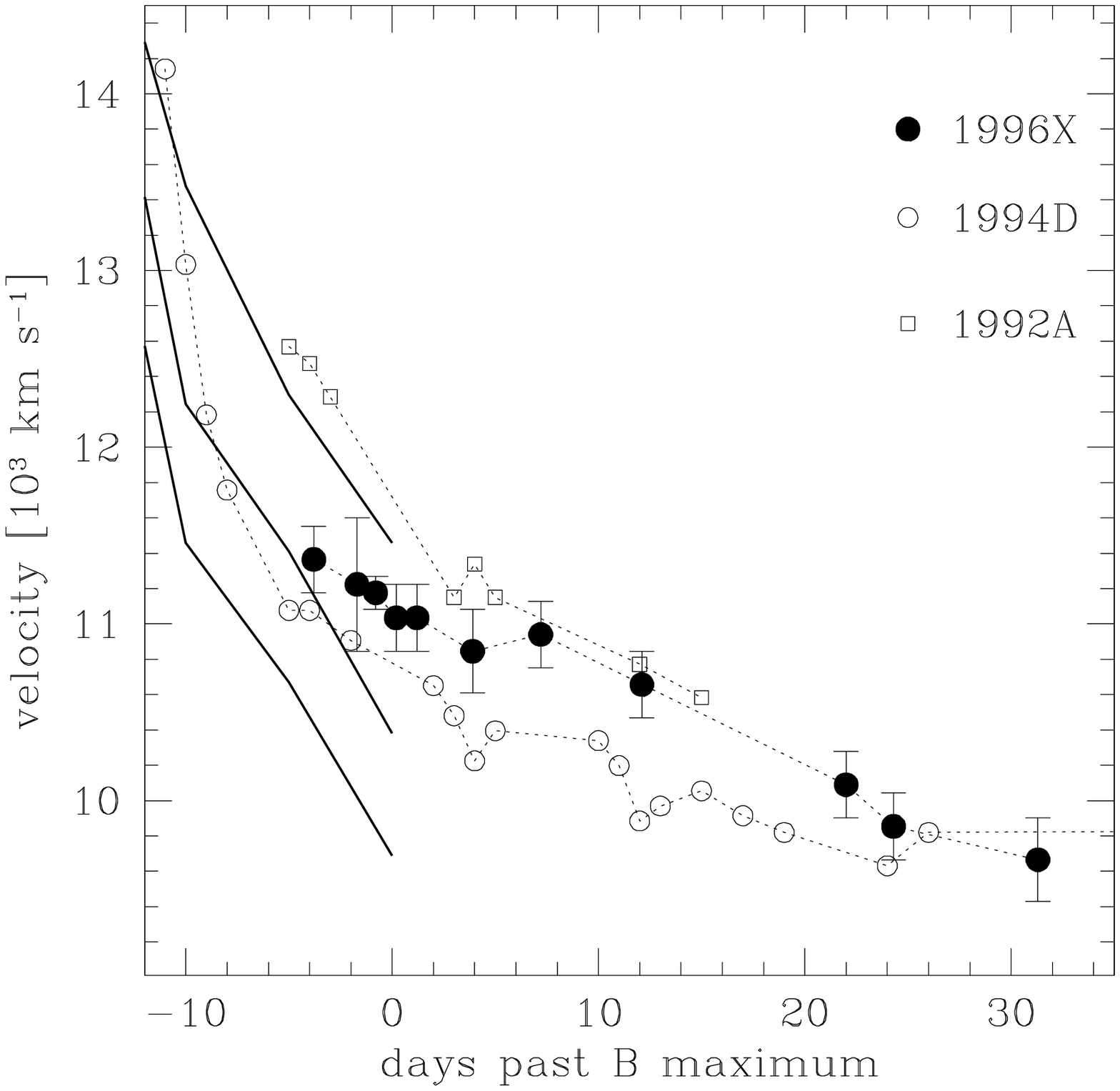,width=0.5\textwidth}
\caption{Comparison of the Si II velocities of SN 1996X, SN 1992A, 
SN 1994D and the model velocities of 10 times normal C+O layer metallicity,
normal and 1/10 normal as defined by Lentz et al., 1999 (dashed
lines, from top to bottom).}\label{vl3}
\end{center}
\end{figure}    

The velocity of the CaII H\&K absorption has a different behaviour. It is very
similar to the velocity of the same line in SN~1994D, indicating an expansion
velocity 2000 km s$^{-1}$ slower than for SN~1992A. This supports the
conclusions of Wells et al. (1994) that $\Delta$ m$_{15}$(B) and the SiII
velocity do not appear to be correlated, while $\Delta$ m$_{15}$(B) and the Ca
II velocity are. This is not easy to explain since CaII H\&K is a stronger 
line than SiII 6355\AA and therefore it is even less indicative of the true
photospheric expansion velocity.

We have also measured in the near-maximum spectra the $R$(Si II) and
$R$(Ca II) ratios introduced by Nugent et al. (1995).  The
values for SN~1996X are 0.25 and 1.39, respectively, very similar to
the values typical for `normal' SNe~Ia, in particular SN 1994D, for
which the values are 0.29 and 1.38, respectively.  According to Nugent
et al. (1995), this indicates that the temperature and
therefore the Ni mass are similar in these two objects.  However, this
may be contradicted by the very different strength of the SiIII line.

Mazzali et al. (1998) included SN 1996X and other SNe Ia in a diagram showing 
a correlation between $\Delta m_{15}$(B) and the FWHM of the 4700 \AA\ nebular
feature, which is a blend of [FeII] and [FeIII] lines, and should thus reflect
the mass and distribution of $^{56}$Ni. Since the preliminary measurements 
used by Mazzali et al. (1998) for their plot are confirmed here, that result 
is also confirmed.
  
\begin{table}    
\footnotesize
\begin{center}
\caption{The measured position,  of the lines of Si II and CaII, 
and their expansion velocity.}
\label{vel}    
\begin{tabular}{cccccc}
\hline
ph$^*$ & Si II & $v_{exp}$(SiII     ) &  Ca II & $v_{exp}$(CaII) \\
   &  [\AA]& [$10^3$ km s$^{-1}$] & [\AA]  & [$10^3$ km s$^{-1}$] \\ 
\hline
  -3.8   & 6158$\pm$4 & 11.4$\pm$.2  &  ---       & --    \\        
  -1.7   & 6161$\pm$8 & 11.2$\pm$.4  & 3824$\pm$3 & 11.8$\pm$.2 \\
  -0.8   & 6162$\pm$2 & 11.2$\pm$.1  & ---        & --    \\       
  +0.2   & 6165$\pm$4 & 11.0$\pm$.2  & 3820$\pm$4 & 12.0$\pm$.3 \\
  +1.2   & 6165$\pm$4 & 11.0$\pm$.2  & 3822$\pm$4 & 11.9$\pm$.3 \\
  +3.9   & 6169$\pm$5 & 10.9$\pm$.2  &  ---       & --    \\       
  +7.2   & 6167$\pm$4 & 10.9$\pm$.2  & 3822$\pm$4 & 11.9$\pm$.3 \\
  +12.1  & 6173$\pm$4 & 10.7$\pm$.2  &  ---       & --    \\       
  +22.0 & 6185$\pm$4 & 10.1$\pm$.2   & 3826$\pm$4 & 11.6$\pm$.3 \\
  +24.3 & 6190$\pm$4 &  9.9$\pm$.2   & 3828$\pm$4 & 11.4$\pm$.3\\ 
  +31.3 & 6194$\pm$5 &  9.7$\pm$.2   & 3829$\pm$4 & 11.2$\pm$.4 \\
  +56.0 &  ---       &       ---     & 3840$\pm$5 & 10.6$\pm$.2 \\
  +57.1 &  ---       &       ---     & 3839$\pm$9 & 10.8$\pm$.5 \\
  +87.6 &  ---       &       ---     & 3841$\pm$8 & 10.4$\pm$.3 \\
\hline                                                       
\end{tabular}                                               
\end{center}

$*$ Relative to the epoch of B maximum JD=2450191.5\\
\end{table}

\section[]{Synthetic spectra}

We fitted three early-time and one late-time spectra with our various codes in
order to gain a deeper understanding of the properties of SN~1996X.

In particular, we modelled the spectra using two different distances.  One set
of models was computed for $\mu=32.02$, which is the value obtained from the
Hamuy et al. \shortcite{hamuy_absmag} calibration of the $M_B - \Delta m_{15}$
relation.  This value makes SN~1996X an average SN~Ia, comparable to SN~1994D,
and somewhat brighter than SN~1992A, even when the larger GCLF distance to this
SN is adopted. In fact, SN~1992 has a slightly larger value of $\Delta m_{15}$
than the other two SNe. Another set of models was computed for the largest
distance, $\mu=32.4$, which was obtained after Phillips et al.
\shortcite{phil99}.  This value makes SN~1996X brighter ($M_B = -19.42$). 
Different distances imply that different parameters are necessary to produce a
synthetic spectrum that fits the observations.

We fitted the early-time spectra using our Monte Carlo code (Mazzali \& Lucy
1993), which has been recently improved to include photon branching (Lucy 1999;
Mazzali 2000). The code is based on the Schuster-Schwarzschild approximation
and uses a large (about 10$^6$) line data base extracted from the line list of
Kurucz \& Bell (1995). Radiative equilibrium is assumed in order to compute the
temperature structure. The epoch, luminosity and photospheric velocity are
input to the code. We used a W7 density structure and mixed composition (Nomoto
et al. 1984). For the lower boundary a black body radiation field is used. Its
characteristic temperature is determined by the temperature iteration, thus
taking into account the effect of backwarming resulting from inward travelling
energy packets (used to sample the radiation field instead of photons) which
are reabsorbed in the photosphere after line absorption and reemission or
electron scattering. Conservation of luminosity above the photosphere ensures a
comparison in flux between synthetic and observed spectra.

\begin{figure*}
\psfig{file=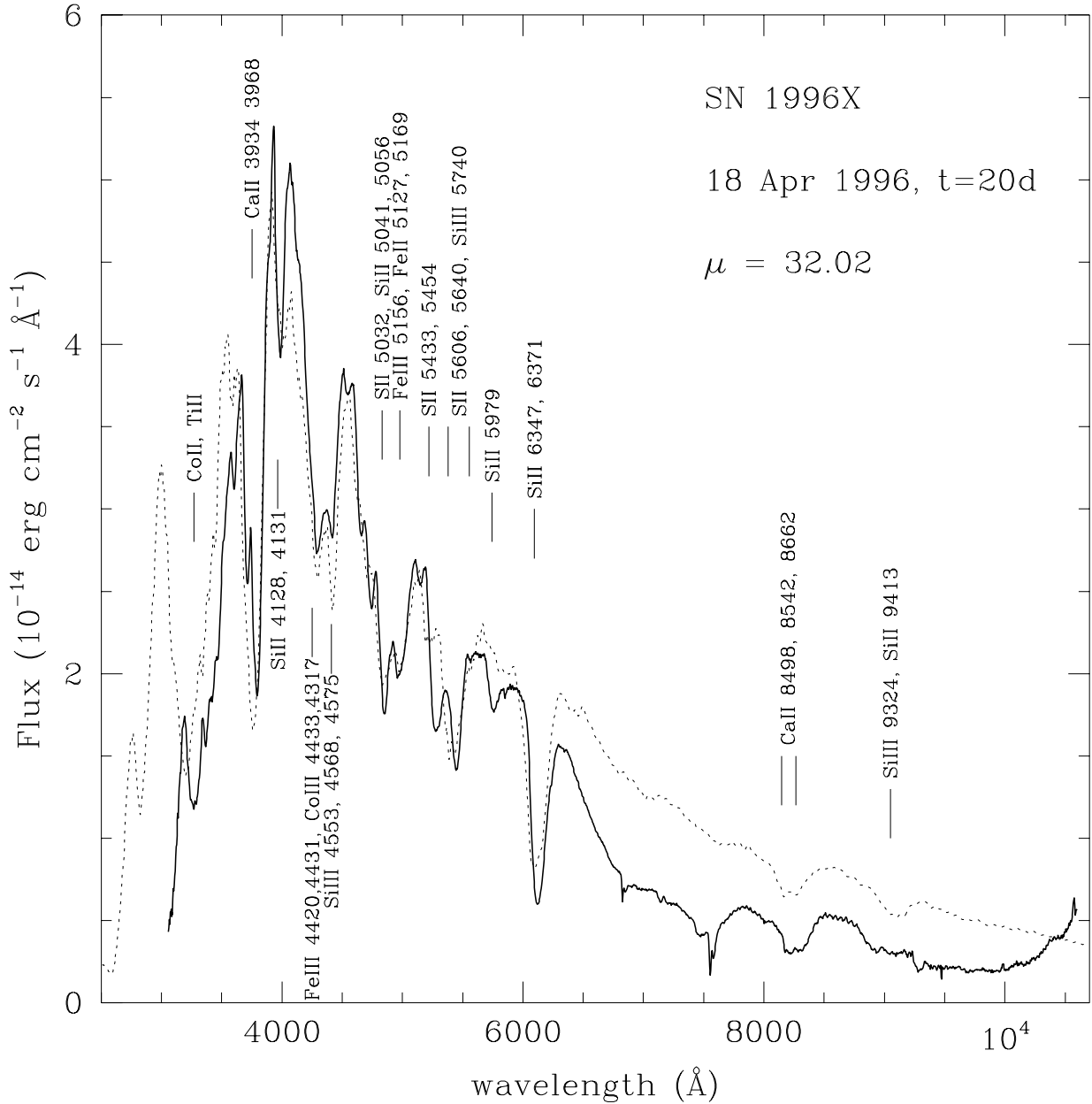,width=\textwidth}
\caption{The observed spectrum of SN~1996X on 18 April 1996, 20 days past 
explosion (thick line) is
compared to a synthetic spectrum based on W7 computed with the input
parameters listed in Table 8 (dotted line).}
\label{fig1}
\end{figure*} 

In order to make our results more sound, we modelled 3 rather widely
spaced epochs: April 18, which is the epoch of $B$ maximum; April 25
and May 10. We adopted $t_B(max) = 20$ days for the delay of B maximum
relative to the epoch of explosion, so our epochs are 20, 27 and 42
days respectively. When fitting a spectrum with our Monte Carlo code,
our input parameters lead to changes in the overall flux ($L$), in the
line velocity ($v_{ph}$) and in the overall nature of the spectrum
(the temperature, which results from both $L$ and $v_{ph}$ and on the
details of the explosion model used). Therefore, given an observed
spectrum, an assumed epoch and distance, we may hope that the
allowed range of the parameters $L$ and $v_{ph}$ can be at least
narrowed down based on the quality of the fit.

The quality of a synthetic spectrum is judged according to its ability to
reproduce the observed spectral features. This is essentially the result of
deriving an appropriate temperature structure, and hence the correct runs of
ionisation and excitation with radius, which is equivalent to velocity in a
homologously expanding medium such as the SN ejecta.

Since we tested distance moduli that are different by 0.4  mag, our two sets of
models have $L$'s that differ by almost a factor of 1.5 in order to reproduce
the observed flux level. If this change in $L$ is not accompanied by a
simultaneous change in $v_{ph}$, the two sets of models will have temperatures
different by a factor $\sim 1.5^{1/4} \sim 1.1$. This would lead to significant
differences in the synthetic spectra. On the other hand, an almost identical
temperature can be obtained if the increase in $L$ by a factor 5 is accompanied
by an increase in $v_{ph}$ by a factor $\sim 1.5^{1/2} \sim 1.2$.  In this case
the temperature of the two models is the same, but the photosphere moves
significantly in radius/velocity space, and thus the line features have
different redshifts: a larger $v_{ph}$ means bluer lines. This is particularly
noticeable for those lines that are reasonably well isolated in wavelength
space.  A second-order effect is also that for a large $v_{ph}$ the amount of
ejecta above the photosphere, with which the energy packets released at the
photosphere can interact, is smaller. This leads to reduced backwarming and to
a flatter temperature structure. If $v_{ph}$ is very large, a given explosion
model may not have sufficient mass above the photosphere to yield the observed
line strengths. All these tests are useful to determine which is the better
value of the distance, at least given a particular explosion model. We now
briefly discuss the models for the three epochs.

\subsection{18 April 1996, day 20 past explosion} 

\begin{figure}
\psfig{file=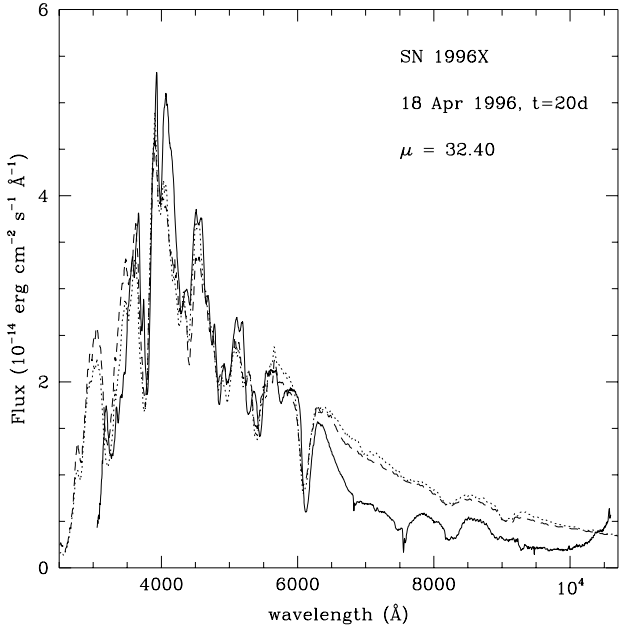,width=0.5\textwidth}
\caption{The observed spectrum of SN~1996X on 18 April 1996 (thick line) is
compared to two synthetic spectra obtained for a large distance: the dotted
line is a spectrum computed using W7 with $v_{ph}=10,000$ km s$^{-1}$; the 
dashed line is a spectrum computed with the same parameters but an {\em ejecta}
mass increased by a factor of 1.6}
\label{fig2}
\end{figure}

The observed spectrum shows signs of a high temperature through the presence of
the Si~III line at $\sim 4400$\AA, the strength of the Si II 6355\AA\ line and
of the two S~II features at 5200 and 5400\AA. For the shorter distance, a good
fit to the spectrum can be obtained for the following parameters: $\log L =
43.05$ (erg s$^{-1}$); $v_{ph} = 8000$ km s$^{-1}$. Other important parameters
and the photometry of the synthetic spectrum are listed in Table 8. The
best-fitting model is shown in Fig.~\ref{fig1}. When the larger distance is
adopted, $L$ increases to 43.24. In order to achieve a reasonable temperature
and to produce the correct excitation/ionisation regime, $v_{ph}$ must increase
to 10000 km s$^{-1}$.  The synthetic spectrum (Fig.~\ref{fig2}, dashed line)
has the correct lines, but all these lines are weaker than in the observed
spectrum. Because of the large $v_{ph}$, the absorbing mass is in fact smaller
than in the model for the shorter distance.  Also, more flux escapes in the
near-UV (below 3200\AA) because line blanketing is not strong enough. Finally,
the velocity of lines such as Si II and S~II is larger than in the observed
spectrum. This shows that the large distance leads to parameters which are less
compatible with the observations. As an instructive example, we also show in
Fig. \ref{fig2} (dotted line) a synthetic spectrum obtained after multiplying
the density by a factor of 1.6: the lines are deeper, but the velocities are
still incorrect. Therefore, the shorter distance appears to be favoured.

\subsection{25 April 1996, day 27 past explosion} 

\begin{figure}
\psfig{file=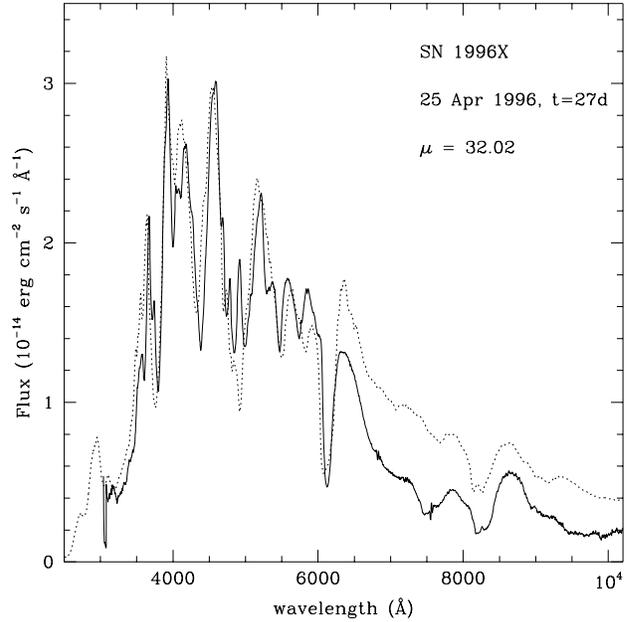,width=0.5\textwidth}
\caption{The observed spectrum of SN~1996X on 25 April 1996 (thick line) is
compared to a synthetic spectrum based on W7 computed with the input parameters
listed in Table 8 (dotted line).}
\label{fig3}
\end{figure} 

The next epoch we modelled, day 27, is marked by a cooling of the
ejecta. This is evident in the disappearance of the Si~III line, the
strengthening of the Si~II and Ca~II lines, and the overall redder
colour of the spectrum. Our short distance model gives a best fit for
$\log L = 42.91$ (erg s$^{-1}$); $v_{ph} = 6750$ km s$^{-1}$. The
temperature (Table 8) is lower than on day 20.  The synthetic spectrum
is shown in Fig.~\ref{fig3}. The poor fit in the red is due to our
assumption of black body radiation at the lower boundary. The lack of
spectral lines in the red means that the effective opacity is much
lower in that region than in the $B$ and $V$ part of the spectrum, and
hence our assumed photospheric radius is larger than its real
value. Still, line features (e.g.  the Ca~II IR triplet) are well
reproduced. Elsewhere, the synthetic spectrum fits the observed one
well. 

\subsection{10 May 1996, day 42 past explosion} 

\begin{figure}
\psfig{file=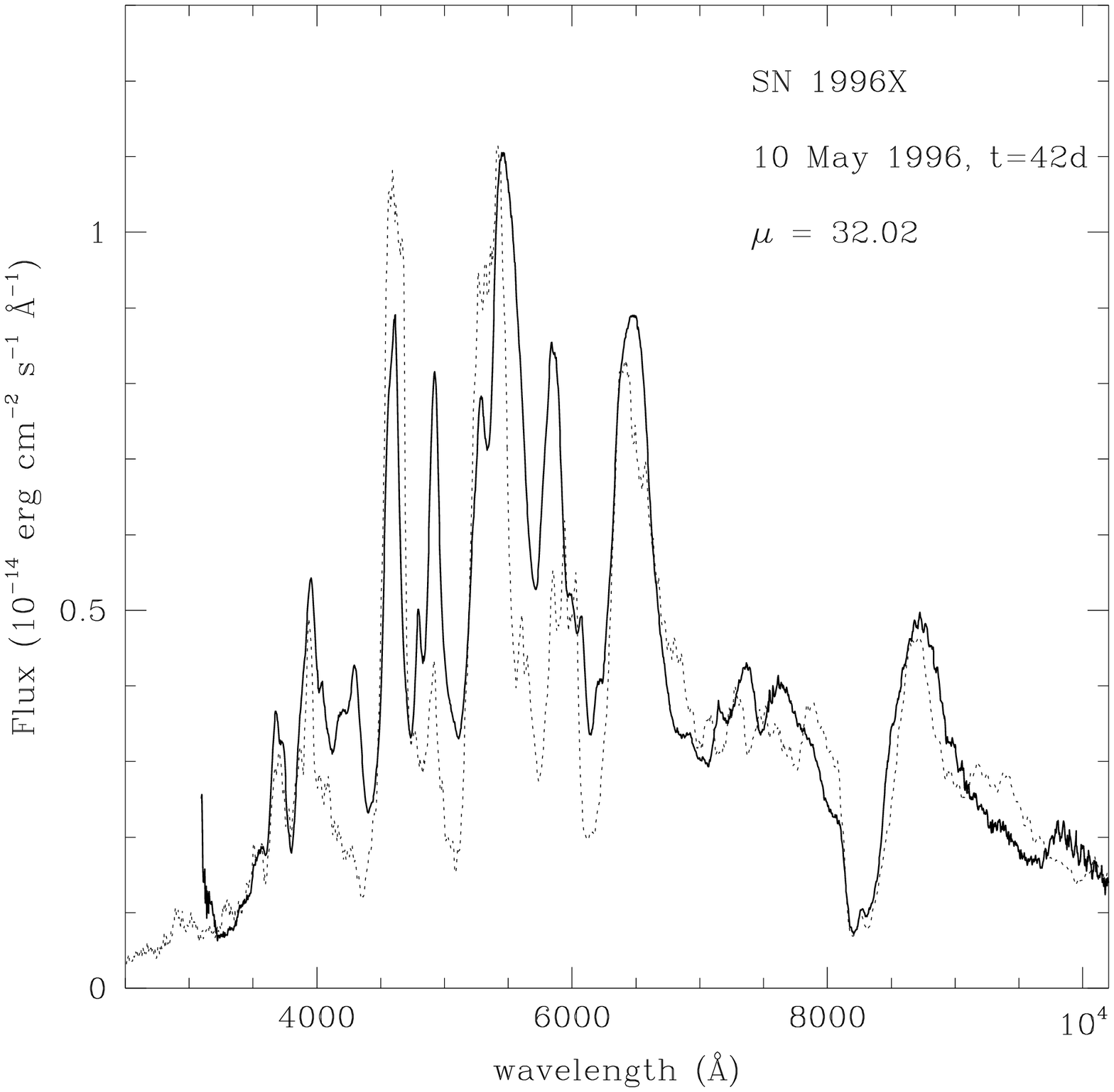,width=0.5\textwidth}
\caption{The observed spectrum of SN~1996X on 10 May 1996 (thick line) is
compared to a synthetic spectrum based on W7 computed with the input parameters
listed in Table 8 (dotted line).}
\label{fig4}
\end{figure}

In the third epoch, the spectrum shows very deep absorptions,
indicating that the photosphere has receded deep inside the
ejecta. The overall colour is very red ($B-V\sim1.2$) and all lines
are from low-ionisation species. Na~I~D appears as a strong line near
5700\AA. Our best fit for the short distance has $\log L = 42.41$ (erg
s$^{-1}$); $v_{ph} = 2500$ km s$^{-1}$, which is very small.  The
temperature is also much smaller than at earlier epochs. The synthetic
spectrum (Fig.~\ref{fig4}) fits the observations reasonably well. The
problem with the red flux is now gone, because the photosphere is so
deep that the relative change of $R_{ph}$ with $\lambda$ is small and
the low temperature means the red flux is relatively high. As is
typical at these epochs, the Na~I~D line could only be obtained by
adding a small amount of neutral Na (Na~I/Na = $10^{-6} r^{-2}$) by
hand. We are not aware of any calculation where the Na~I~D line is
correctly reproduced at this advanced epoch (see e.g. the NLTE
synthetic spectrum shown in Pauldrach et al. 1996).

In this case the absolute change of $v_{ph}$ is small, and the
overlying mass is large in both cases, so that they cannot be
distinguished so easily. Also, the run of ionisation with radius is
rather flat, because of the low temperature, and thus no line really
marks $v_{ph}$, as did e.g. S~II at earlier epochs.

At more advanced epochs the spectrum becomes contaminated by nebular emission
and cannot be successfully modelled with our Monte Carlo code. Note that the
$L's$ used at the three epochs are consistent with the values of
$L_{Bol}$ derived from the observations in Fig.~\ref{bol}.

\subsection{10 Feb 1997, day 318} 

We used a 1-zone NLTE code to compute the nebular spectrum at 10 months. The
code computes the instantaneous deposition of $\gamma$-rays and positrons from
the decay of $^{56}$Co and uses this as the heat input into the SN nebula,
which is bounded by an outer velocity and is composed mostly of Fe at this late
epoch. The outer nebular velocity is determined fitting the width of the
emission lines; the temperature and density, and hence the mass of $^{56}$Ni
originally synthesized, by fitting the ratio of the two strong lines in the
blue. The bluer of these lines is in fact a blend of mostly Fe~III] lines,
while the redder one contains about equal contributions of Fe~II] and Fe~III]
lines.

\begin{figure}
\psfig{file=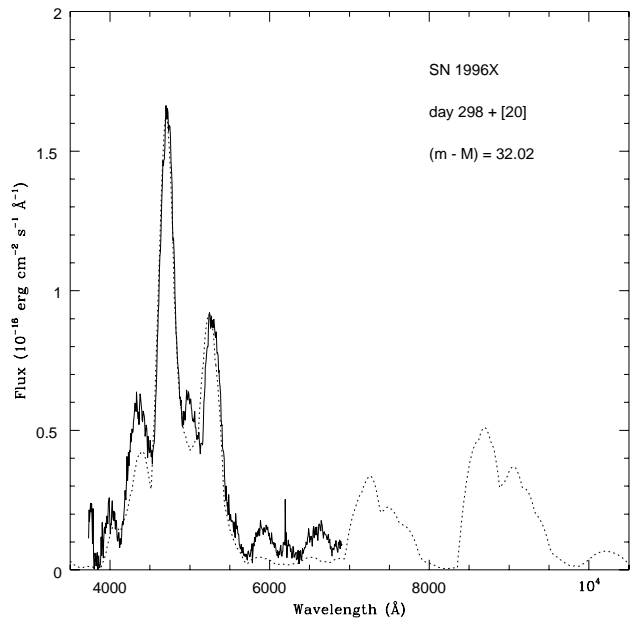,width=0.5\textwidth}
\caption{The observed spectrum of SN~1996X on 10 Feb 1997 (thick line) is
compared to a synthetic NLTE nebular spectrum computed using $M(Ni)=0.61$
M$_{\odot}$ and $v = 9250$ km s$^{-1}$ (dotted line). }
\label{fig5}
\end{figure}

The spectrum can be fitted equally well for both distances if different
$^{56}$Ni masses are used. The outer velocity is 9250 km s$^{-1}$. For the
short distance a good fit is obtained for $M(Ni) = 0.65$ M$_{\odot}$ and a
total mass in the nebula (i.e. within the outer velocity) of 0.70 M$_{\odot}$. 
This can be compared with $M(Ni) = 0.6$ M$_{\odot}$ and a total mass of 0.85
M$_{\odot}$ contained within 9500 km s$^{-1}$ in W7. The fit is shown in
Fig.~\ref{fig5}. These values indicate that SN~1996X can be regarded as a
'normal' SN~Ia.

If on the other hand the long distance is used, a good fit requires $M(Ni) =
0.88$ M$_{\odot}$ and a total mass in the nebula within 9500 km s$^{-1}$ of 
1.00 M$_{\odot}$. This values are larger than in W7, and may lead to a total 
ejected mass larger than the Chandrasekhar limit if a density distibution
similar to that of W7 were adopted at velocities larger than the outer nebular
velocity. Clearly, the estimate of the distance has consequences on the
interpretation of the properties of SN~1996X.

\begin{table*}
\begin{center}
\caption{Early-time, short distance models for SN 1996X.}
\label{mod}
\begin{tabular}{|c|c|c|c|c|c|c|c|c|c|c|c|c|c|}  
\hline
Date & epoch & $L$ & $v_{ph}$ & T$_{eff}$ & T$_{BB}$ & $\rho(ph)$ & M(ab) & W & $U$ & $B$ & $V$ & $R$ & M$_{Bol}$\\
     &   d & log erg/s & km/s &    K      &    K     & g/cm3      & M$\odot$  &\multicolumn{5}{c}{synthetic}   \\
\hline
 18/04/96 & 20 & 43.05 & 8000 & 9500 & 12000 & -13.35 & 0.70 & 0.50 & 12.96 & 13.30 & 13.21 & 12.85 & -18.90 \\
 25/04/96 & 27 & 42.91 & 6750 & 8200 &  9700 & -13.54 & 0.84 & 0.36 & 13.70 & 13.70 & 13.35 & 13.11 & -18.55 \\ 
 10/05/96 & 42 & 42.41 & 2500 & 8200 & 10000 & -13.40 & 1.25 & 0.41 & 15.82 & 15.67 & 14.38 & 14.10 & -17.30 \\
\hline
\end{tabular}
\end{center}
\end{table*}        

\section[]{Conclusions}

We have presented photometric and spectroscopic data of Supernova
1996X obtained at ESO-La Silla and Asiago.

By examining the UBVRIJ light curves, the colour curves and the
spectra, we conclude that SN 1996X is a `normal' SN Ia, i.e. it does 
not show peculiarities resembling `overluminous' or 'underluminous'
SNe~Ia. We measure $\Delta m(15)$(B) = 1.31$\pm$0.08 mag, similar to
the values found for the `normal' SNe~Ia 1992A and 1994D, the most
similar to SN 1996X also in their spectroscopic evolution.

Different absolute magnitudes are obtained if one adopts the distance
modulus for the host galaxy obtained by Faber et al. (1989) using the
D$_n-\sigma$ relation or that derived from the distance-redshift relation.

The value of the interstellar extinction A$_B = 0.30$ reported by Schlegel et
al. (1998) is consistent with the value we find using the relation between the
equivalent width of the Na I D line in the spectra and the interstellar
extinction. However, both the {\em Multicolour Light-curve Shape} and the {\em
snapshot} methods by Riess et al. (1996, 1998a) yield as a result that the
absorption is zero. The problem is that the relations between light curve shape
and colour and the absolute magnitude of a SN~Ia are not yet fully reliable,
probably because of problems in estimating distance and reddening in the sample
used for the calibration. The same problems may explain the discrepancy between
different calibration of the $M_B - \Delta m_B(15)$ relation
\cite{hamuy_absmag,phil99}.

We have modelled four spectra at various epochs, using our Monte Carlo code and
a 1-zone NLTE nebular code, to investigate the problems related to determining
the distance. We found that a large distance is difficult to accomodate.  If we
use the distance obtained from Hamuy et al. \shortcite{hamuy_absmag}
calibration of the $M_B - \Delta m_B(15)$ relation, $\mu=32.02$, reasonable
synthetic spectra for the photospheric-epoch can be obtained using a standard
W7 density structure. The late-time spectra can be fitted using $M(Ni) = 0.65$
M$_{\odot}$, which is appropriate for W7, and a similar value is also necessary
to reproduce the light curve near maximum with a Monte Carlo code. However, if
we use a larger distance, $\mu=32.4$, the early-time spectra are better 
reproduced for a photospheric velocity larger than the observed one, the
late-time spectra require a larger $^{56}$Ni mass, and models in general models
favour an ejecta mass larger than the Chandrasekhar mass.  
For the shorter $D_N - \sigma$ distance, the opposite problem arises.

One of the main problems in our understanding the properties of SN~Ia is that
we do not know enough yet about the progenitors and the explosion mechanism of
SN~Ia. Objects with similar $\Delta m_{15}$(B) and general spectral behaviour
could have different evolution and intrinsic characteristics, in particular
different expansion velocities, as is shown in Figure 11, and perhaps also
different absolute magnitudes.

We need to get more high quality data for nearby SN~Ia and their host
galaxies to clarify this, as SN~Ia are now the most reliable distance
indicators on a cosmological scale, and the value of their absolute magnitude
has to be calibrated using nearby objects.

Since the host galaxy of SN~1996X is an elliptical galaxy, the Globular
Cluster Luminosity Function (GCLF) method (Drenkahn \& Richtler, 1999, Della
Valle et al., 1998) could provide a velocity-independent estimate of its
distance.

\section*{Acknowledgments}

We thank all the astronomers who kindly observed SN 1996X for us at
ESO-La Silla: Chevallier, L., Tappert, C., Metanomski, A.D.F.,
Massone, G., Wickmann, L.A., Van Bemmel, I., Spyromilio, J., Zanin,
C..  We especially thank Fouque, P., who provided reduced IJK data of
the SN obtained at the DENIS 1m telescope.

\end{document}